\documentclass[a4paper,fleqn]{cas-sc}
\usepackage{longtable}
\usepackage{float}
\usepackage{placeins}
\usepackage{hyperref}
\usepackage{enumitem} 
\usepackage{setspace}
\usepackage{algorithm}
\usepackage{algpseudocode}
\usepackage{mdframed}
\usepackage{graphicx}%
\usepackage{afterpage}
\usepackage[numbers]{natbib}
\usepackage{amsmath}
\usepackage[justification=centering, width=1.5\textwidth]{caption}
\usepackage{lineno}
\usepackage{multirow}
\usepackage[table,xcdraw]{xcolor}
\definecolor{ForestGreen}{RGB}{34,139,34}
\definecolor{BrickRed}{RGB}{192,0,0}
\definecolor{RoyalBlue}{RGB}{65,105,225}

\begin{document}
\let\WriteBookmarks\relax
\def\floatpagepagefraction{1}
\def\textpagefraction{.001}
\shortauthors{Khawale  et~al.}
\shorttitle{Rapid Kirigami Simulation using the Bar \& Hinge Approach}  

\title [mode = title]{Rapid Kirigami Simulation using the Bar \& Hinge Approach}


\author[1]{Raj Pradip Khawale}  
\cormark[1] 
\ead{E-mail address: rkhawal@clemson.edu}
\cortext[cor1]{Corresponding author}   

\address[1]{Department of Civil and Environmental Engineering, University of Michigan, Ann Arbor, MI 48109, USA}




\author[2]{Elaheh Mehdizadeh}
\author[2]{John Brigham}

\address[2]{Department of Civil and Environmental Engineering, 
University of Pittsburgh, Pittsburgh, 15261, PA, United States}


\author[1]
{Evgueni T. Filipov}



\begin{abstract}
Kirigami, the art of cutting sheets, offers unique characteristics such as shape morphing, stretchability, and adaptability, with applications in deployable and reconfigurable structures. 
While Finite Element (FE) approaches are widely used to analyze kirigami structures, they are computationally intensive and prone to convergence issues. On the other hand, existing simplified and theoretical models for evaluating kirigami are typically restricted to specific designs and analytical scenarios.
This paper proposes a generalized, computationally efficient reduced-order model based on the bar and hinge approach for analyzing any kirigami system. The model represents the entire kirigami structure using truss bars, bending hinges, and torsional springs, capturing both the deformation and internal forces of the system.
We derive stiffness expressions for kirigami structures that experience stretching and out-of-plane bending deformations, as well as for structures that experience only in-plane stretching and rotations.
The accuracy and robustness of the model are validated through comparisons with FE analysis and experimental data on both single-cut kirigami designs and more complex designs with multiple cuts. Our results demonstrate less than $5\%$ difference in deformation predictions and achieve at least a tenfold computational speedup compared to FE simulations. 
Additionally, we demonstrate the model’s capabilities through simulating complex scenarios, including kirigami-skinned crawlers, high-throughput property-space exploration, and in-plane kirigami-inspired metamaterials. These examples demonstrate computational capabilities that enable large-deformation structural simulations and high-throughput design exploration beyond the practical limits of conventional FE methods.


\end{abstract}


\begin{highlights}
\item Reduced-order generalized bar and hinge model for rapid simulation of kirigami systems
\item Introduces three-node torsional spring for rotational resistance without rotational DOFs
\item  Validated with <5\% shape and <10\% stiffness difference versus FE models and experiments
\item Achieves ~10× speedup over conventional finite element model simulations
\item Demonstrates simulations of kirigami-skinned crawlers and large-scale systems
\end{highlights}


\maketitle



\section{Introduction}

Kirigami—derived from the Japanese words for “cut” (kiri) and “paper” (gami)—is a design paradigm in which sheets are precisely cut and often folded to create complex, transformable two- or three-dimensional configurations from initially flat, sheets \cite{zhai2021mechanical, ning2018assembly}. In recent years, the concept has expanded far beyond artistic paper crafting into engineering and materials science \cite{sun2021geometric,bu2026hybrid}. By introducing tailored cuts, planar sheets can undergo large deformations \cite{hong2024ultra}, shape morphing \cite{cheng2020kirigami,ying2025inverse}, tunable stiffness \cite{hwang2018tunable}, auxetic responses \cite{du2023auxetic, khawale2024tiling}, energy absorption capabilities \cite{he2026geometrically}, and reconfigurability \cite{zheng2022kirigami}. These capabilities enable applications in flexible electronics \cite{won2019stretchable,tang2025design}, soft robotics \cite{sedal2020design}, deployable structures \cite{wang2022kirigami}, lightweight architected materials \cite{zhang2022kirigami}, metamaterials \cite{tang2017programmable, khawale2024efficient,khawale2025aperiodic, meeussen2025textile}, building facades \cite{arauz2024evaluation}, and biomedical devices \cite{wu2025origami}. For example, kirigami-inspired plate–lattice structures exhibit ultralight weight yet high strength, making them promising for aerospace and automotive systems \cite{pan2026coupled}.
Given this broad potential, computational simulators for kirigami systems are increasingly necessary. However, the inherent complexities of kirigami—such as large-deformation kinematics, nonlinear buckling, thin-sheet behavior, and coupling between geometric patterning and mechanics—make it challenging for any single simulation framework to be robust, generalizable, and computationally efficient. As a result, design exploration remains laborious, and engineering optimization of kirigami structures is severely constrained.
%

Kirigami structures are typically modeled using finite element (FE) simulations with shell elements, which provide detailed stress and deformation fields but require extensive effort in geometry setup, computation, and post-processing \cite{khalilzadehtabrizi2025analysis, gomes2025design,khawale2019finite}. Despite their accuracy, shell formulations can be computationally demanding and susceptible to numerical issues such as shear and membrane locking \cite{kumar2024deep,khawale2023efficient}. As the sheet thickness becomes small, additional stabilization or enhanced formulations are often needed to maintain accuracy, and localized instabilities can further hinder global convergence. However, in many applications the primary objective is to capture the overall deformation response rather than fine-scale stress variations. For such analyses, a simplified yet robust modeling approach is preferred—one that can efficiently track the global deformations of kirigami structures while remaining less sensitive to local instabilities.

Apart from FE analysis, the simulation of kirigami-inspired structures has advanced along two complementary directions. First, analytical models—such as those by Zheng et al. \cite{zheng2023modelling}—capture effective, cell-averaged kinematics and the mechanics associated with panel rotations, hinge bending, and inter-panel interactions, enabling efficient physics-based prediction of large deformations with good agreement to experiments. Similarly, Chen et al. \cite{chen2025kirigami} developed a theoretical framework based on energy conservation and micro-element discretization to characterize the kinematics and electromechanical coupling behavior of kirigami-inspired substrates. Despite their computational efficiency, such analytical approaches are often problem-specific and difficult to generalize across diverse geometries and loading conditions, and deriving closed-form relationships for structural performance remains challenging.

Second, data-driven surrogate models—including recent GAN-based full-field predictors \cite{xiang4809373gan}—have been developed to learn mappings from cut patterns and loading conditions to deformation and stress fields, enabling orders-of-magnitude speedups for rapid design iteration and inverse analysis. Zhang et al. \cite{zhang2025deep} further proposed a deep learning framework to capture the nonlinear relationship between kirigami geometric parameters and their mechanical responses. While these models significantly accelerate predictions and reduce reliance on exhaustive FE simulations, their performance is inherently constrained by the availability and diversity of training datasets, limiting their generalizability beyond the specific problem domains on which they are trained.

In addition to computational approaches, kirigami behavior has been extensively investigated through systematic experimental studies. These typically involve fabrication of physical specimens and controlled laboratory testing, which not only validate theoretical and numerical models but also reveal new mechanical phenomena. Representative efforts include tensile characterization of 3D-printed flexible kirigami structures \cite{nakajima2020experimental}, mechanically driven 3D assemblies \cite{ning2018assembly}, studies of rotational kirigami systems \cite{de2023experimental}, evaluation of rigidity and failure in non-uniformly deforming kirigami \cite{taniyama2019design}, and validation of auxetic metamaterial models \cite{kim2025design}. Although experimental methods provide the most reliable characterization of mechanical behavior, they are resource-intensive and impractical for large-scale or iterative design exploration.

Despite these advances, existing approaches for simulating kirigami-inspired structures exhibit notable limitations. FE analysis provides high fidelity and general applicability; however, it is computationally demanding and can encounter convergence difficulties under large, highly nonlinear deformations. In contrast, analytical and data-driven approaches offer improved efficiency but are often problem-specific and lack robustness across varying geometries and loading conditions. Consequently, there remains a need for a simulation framework that is computationally efficient, generally applicable, numerically robust, and capable of accurately capturing the global deformation behavior of kirigami systems.


To address this need, this work presents a bar and hinge-based framework for the simulation of kirigami structures. This approach represents the kirigami sheet as a discrete system of truss bars and bending hinges instead of continuum shell elements. This representation significantly reduces the number of degrees of freedom and eliminates the repeated stress evaluations and numerical integration required in conventional FE formulations, resulting in a computationally efficient model. Furthermore, the formulation depends only on the nodal coordinates and bar connectivity and employs simplified mechanical relationships, making the reduced-order model sufficiently general and robust for complex large-deformation analyses. Owing to its generality and computational efficiency, the proposed framework can be readily applied to a wide range of kirigami patterns, enabling parametric studies and design exploration. In addition to nonlinear structural analysis, the framework can also be used for eigenvalue analysis to characterize global deformation modes and stiffness.


Over the past decade, the bar and hinge framework has been progressively refined to improve its accuracy and applicability in modeling origami-inspired structures. Early formulations represented fold lines as rotational springs and panels as bar elements, enabling the capture of folding, bending, stretching, and shearing deformations. The N4B5 model \cite{schenk2011origami,liu2017nonlinear} captured diagonal bending but exhibited anisotropic in-plane behavior, which was later improved by the N5B8 model \cite{filipov2017bar} through additional nodes and bars, allowing bending along both diagonals. Subsequent developments extended these models to arbitrary polygonal panels and introduced compliant crease formulations to account for extensional and torsional effects \cite{zhu2019simulating,zhu2020bar}. More recently, the framework has been further expanded to model curved crease origami \cite{woodruff2020bar} and dynamic behavior by incorporating nodal masses \cite{xia2021deployment,xia2019dynamics,dong2021dynamic}.

Building on these developments, the present work extends the bar and hinge formulation to kirigami systems. The discretization employs bars and hinges for three-dimensional analysis, and bars with torsional springs for in-plane modeling. Bars represent in-plane deformation, hinges capture out-of-plane bending, and torsional springs provide rotational stiffness. This formulation accommodates a range of geometries, boundary conditions, and material properties while maintaining computational efficiency. The formulation is developed to capture deformation features characteristic of kirigami, including large deformations, out-of-plane buckling, and in-plane rotational mechanisms.

There are several important contributions of this paper, which can be summarized as follows:
\begin{itemize}
    \item A computationally efficient, generalizable, and easy-to-use reduced-order model is introduced for kirigami simulation using the bar and hinge approach. The complete MATLAB implementation is publicly available in an open-source GitHub repository \href{https://github.com/Rajkhawale/KirigamiAnalysis-ReducedOrderModel}{[GitHub repository]}.
    \item A three-node torsional spring element is introduced to model rotational resistance in in-plane kirigami simulations, eliminating explicit rotational degrees of freedom while maintaining a compact and computationally efficient formulation.
    \item The deformation behavior and mechanical responses are compared with results from experiments and commercial software, showing less than $5\%$ and $10\%$ differences in deformed shapes and stiffness values, respectively.  
    \item This approach demonstrates nearly a tenfold improvement in computational efficiency compared with the conventional FE approach.  
    \item The advantages and applicability of this model are illustrated through a parametric study for property space exploration, complex analyses of kirigami-skinned crawlers, and efficient two-dimensional metamaterial analyses.  
    
\end{itemize}

The reminder of the paper is organized as follows. Section 2 provides stiffness derivation for kirigami systems using bar and hinge descretization. Section 3 provides a discretization approach, necessary boundary conditions, and a nonlinear analysis solution scheme. Section 4 details the comparative results with experimental and commercial FE software. Applicability and benefits of the proposed bar and hinge model is demonstrated in Section 5 through some computationally intensive problems. Section 6 presents advantages and limitations of the proposed method and finally, Section 7 summarizes conclusions and future work.

\section{Bar and hinge approach for kirigami systems}


 \begin{figure*}
    \centering
    \includegraphics[width=0.99\textwidth]{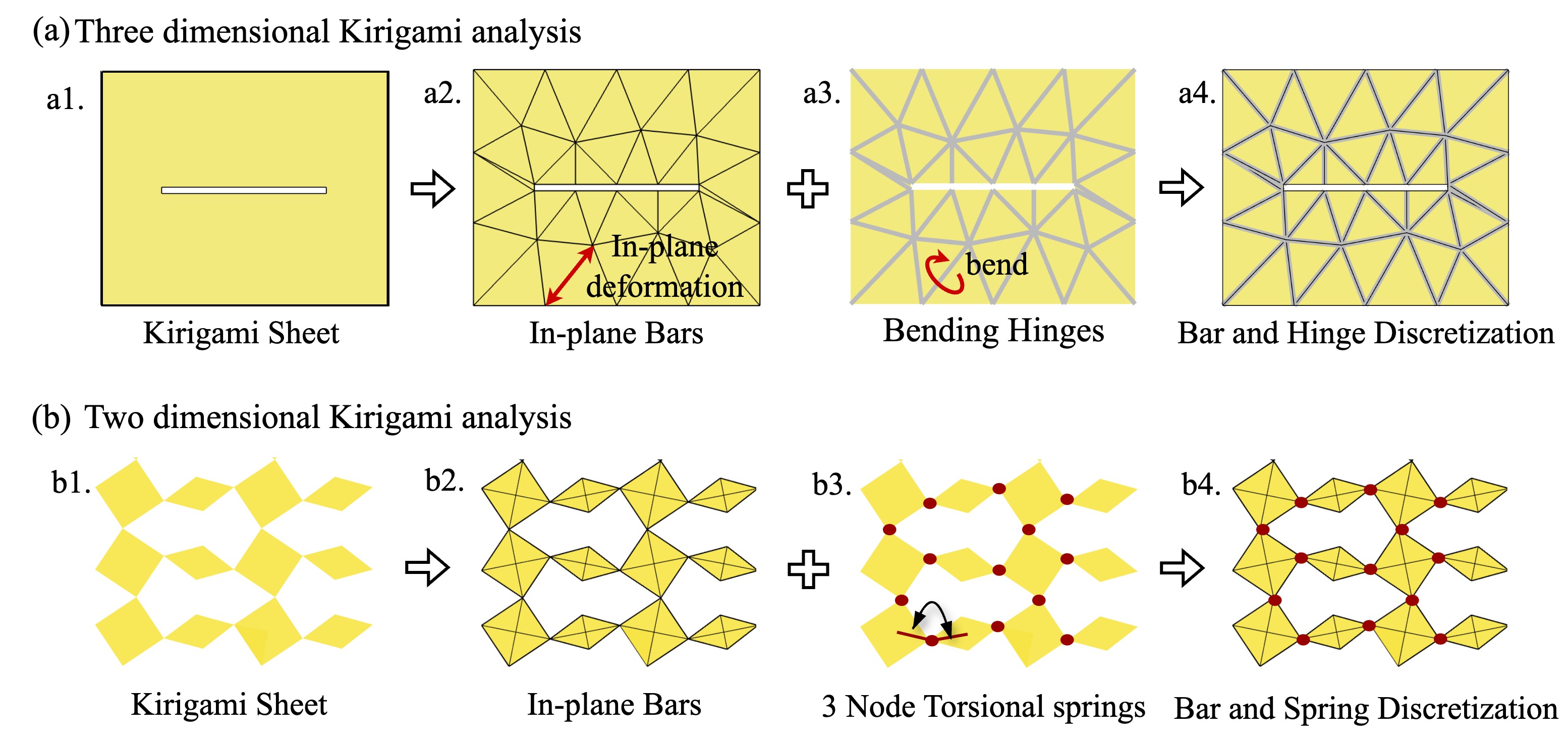}
    \caption{Representation of the bar and hinge model for a kirigami structure. (a) Representation for a three dimensional kirigami analysis for a rectangular structure with a single cut. Each black line represents a bar for in-plane deformation. Each thick gray line represents a bending hinge. Two triangles can bend along their common edge, as shown by the red marking. (b) Representation for a two dimensional kirigami analysis. Each black line represents a bar for in-plane deformation. Each red dot denotes node torsional spring. Two red bars can rotate along the common red dot, as shown in (3).}
    \label{Bar_&_hinge_discretization}
\end{figure*}

The bar and hinge approach is a reduced-order structural mechanics framework for modeling the deformation and internal force response of thin structures. A schematic representation of the proposed framework for both three-dimensional and two-dimensional kirigami analyses is shown in Figure~\ref{Bar_&_hinge_discretization}. In this approach, a kirigami structure is discretized into three types of elements: in-plane bars, bending hinges, and torsional springs. The in-plane bars are connected at nodes and carry loads only along their axial direction, thereby representing the in-plane stiffness of the structure, including its global axial and shear responses (Figure~\ref{Bar_&_hinge_discretization}(a2) and (b2)).
Bending hinges are incorporated only in the three-dimensional model and are represented as rotational springs along the shared edges between adjacent triangular facets. These elements capture out-of-plane bending and provide the flexural stiffness of the structure (Figure~\ref{Bar_&_hinge_discretization}(a3)). In contrast, three-node torsional spring (3-NTS) elements are employed only in the two-dimensional model to capture rotational effects within the plane. The 3-NTS elements are placed at the common node shared by two adjacent quadrilateral elements, as shown in Figure~\ref{Bar_&_hinge_discretization}(b3).

Overall, the 3D model shown in Figure~\ref{Bar_&_hinge_discretization}(a) explicitly captures out-of-plane bending and buckling through bending hinge elements. In contrast, the 2D model shown in Figure~\ref{Bar_&_hinge_discretization}(b) is specifically formulated for kirigami structures undergoing strictly in-plane deformation and, therefore, does not account for out-of-plane effects.
From a physical perspective, the kirigami sheet is represented as a network of translational and rotational springs. Specifically, the truss bars act as axial springs, the bending hinges act as rotational springs for out-of-plane bending, and the 3-NTS elements act as rotational springs for in-plane deformation. Together, these elements provide an efficient reduced-order representation of both in-plane and out-of-plane deformation behaviors in kirigami structures.

 The following subsections detail the stiffness evaluation for in-plane bars, bending hinges, and rotational spring elements and then the methods to combine them and obtain the global stiffness matrix for the analysis. 


\subsection{Global stiffness derivation}

The formulation of a bar and hinge model  follows a matrix-based structural analysis framework. As shown in Figure~\ref{Bar_&_hinge_discretization}, for three-dimensional analysis, the stiffness contributions from bars and bending hinges are considered. For two-dimensional analysis, the stiffness contributions from bars and rotational spring elements are assembled additively into a global stiffness matrix. Each node has three translational degrees of freedom in three-dimensional analysis and two in two-dimensional analysis, resulting in global stiffness matrices of size $3n\times3n$ and $2n\times2n$, respectively, where $n$ is the total number of nodes.This formulation avoids the introduction of rotational degrees of freedom and is fully compatible with standard matrix structural analysis solvers.

The previously established bar and hinge model by \cite{schenk2011origami,filipov2017bar} is improved and extended in this work for kirigami. The total stiffness matrix for the three and two-dimensional kirigami analysis is constructed as follows: 
\vspace{-1mm}
\begin{equation} \label{K_global_3D}
\mathbf{K}_{3D} =
\begin{bmatrix} \mathbf{C} \\ \mathbf{J}_{B} \end{bmatrix}^{T}
\begin{bmatrix}
\mathbf{D}_{S} & 0 \\
0 & \mathbf{D}_{B}
\end{bmatrix}
\begin{bmatrix} \mathbf{C} \\ \mathbf{J}_{B} \end{bmatrix},
\end{equation}

\vspace{-5mm}
\begin{equation} \label{K_global_2D}
\mathbf{K}_{2D} =
\begin{bmatrix} \mathbf{C} \\ \mathbf{J}_{R} \end{bmatrix}^{T}
\begin{bmatrix}
\mathbf{D}_{S} & 0 \\
0 & \mathbf{D}_{R}
\end{bmatrix}
\begin{bmatrix} \mathbf{C} \\ \mathbf{J}_{R} \end{bmatrix},
\end{equation}

\noindent where $\mathbf{C}$ is the bar compatibility matrix, $\mathbf{J}_B$ is the Jacobian matrix associated with discrete bending hinges, and $\mathbf{J}_R$ is the Jacobian matrix corresponding to three-node torsional spring elements. The diagonal matrices $\mathbf{D}_S$, $\mathbf{D}_B$, and $\mathbf{D}_R$ store the constitutive stiffness parameters for bars, bending hinges, and three-node torsional springs, respectively. 

Eqs. \ref{K_global_2D} and \ref{K_global_3D} may be equivalently written as
\vspace{-1mm}
\begin{equation}
   \mathbf{K}_{3D} = \mathbf{C}^T\mathbf{D}_S\mathbf{C} + \mathbf{J}_B^T\mathbf{D}_B\mathbf{J}_B = \mathbf{K}_S + \mathbf{K}_B,
\end{equation}

\vspace{-6mm}
\begin{equation}
  \mathbf{K}_{2D} = \mathbf{C}^T\mathbf{D}_S\mathbf{C} + \mathbf{J}_R^T\mathbf{D}_R\mathbf{J}_R = \mathbf{K}_S + \mathbf{K}_R,
\end{equation}

\noindent where, $\mathbf{K}_S$, $\mathbf{K}_B$, and $\mathbf{K}_R$ denote the stiffness contributions from bars, bending hinges, and three-node torsional spring elements, respectively. All notations have also been added to the nomenclature section of the Appendix. This formulation highlights the additive contributions of bar stretching and shearing, sheet bending, and joint rotation to the global stiffness matrix.

 \begin{figure*}
    \centering
    \includegraphics[width=0.95\textwidth]{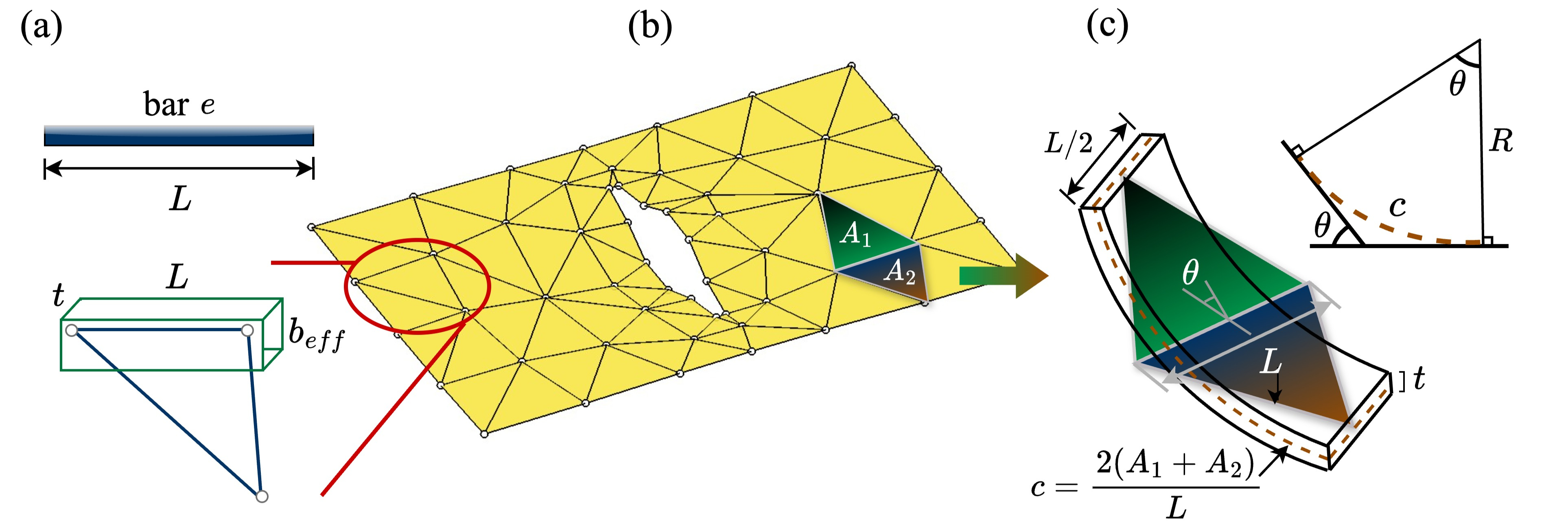}
    \caption{Bar and hinge element definition. (a) Representation of a bar with a cuboid. $L$ and $b_{eff}$ are the length and the breadth of the bar, respectively, and $t$ is the thickness of the structure. (b) Discretized kirigami structure. (c) Hinge representation with continuous curvature. }
    \label{Hinge_description}
\end{figure*}

\subsection{In-plane bar stiffness}

The bars in the kirigami bar and hinge model capture in-plane deformations of the structure, including axial stretching and transverse shearing induced by global loading and constraints. The material properties of the structure, the geometric dimensions, and the discretization of the mesh are incorporated into the bar stiffness definition. To identify the in-plane stiffness, the continuous structure is decomposed into a frame structure, and each individual element is modeled as a bar, as shown in Figure~\ref{Hinge_description}a.  

The stiffness of each bar element is expressed as

\vspace{-3mm}
\begin{equation} \label{eq:bar_stiffness}
\begin{aligned}
k_{s} &= F_s \frac{E A_{\text{eff}}}{L}, & 
A_{\text{eff}} = \frac{A_T t}{L},
\end{aligned}
\end{equation}

\noindent where $E$ is the Young’s modulus, $L$ is the bar length, and $A_{\text{eff}}$ is an effective cross-sectional area representing the portion of the structure associated with the bar. Here, $A_T$ denotes the area of the associated triangular facet, and $t$ is the structure thickness. The effective cross-sectional area of the bar, $A_{\text{eff}}$, is obtained by computing the effective bar width $A_T/L$ and multiplying it by thickness. This gives us lumped cross-sectional area of the structure into the bar.
$F_s$ is a scaling factor obtained through a calibration procedure using a representative triangular panel subjected separately to uniaxial tension and in-plane shear. The corresponding displacements predicted by the discrete bar and hinge model were compared with those of an equivalent plane-stress continuum model, and $F_s$ was iteratively adjusted to minimize the deviation between the two responses. The optimal value of $F_s$ was found to be 0.85, and this value is used throughout the present work. For our kirigami model, this factor primarily regulates the magnitude of force required to initiate out-of-plane buckling.

The bar stiffnesses are assembled into the diagonal matrix $\mathbf{D}_S$, and their contribution to the global stiffness matrix is obtained through the compatibility relation

\vspace{-3mm}
\begin{equation} 
    \mathbf{e}_S = \mathbf{C} \mathbf{u},
\end{equation}

\noindent where $\mathbf{e}_S$ is the vector of bar extensions and $\mathbf{u}$ is the global nodal displacement vector.

\subsection{Bending hinge stiffness}

Kirigami structures exhibit complex mechanical behavior due to pronounced out-of-plane bending and buckling \cite{rafsanjani2017buckling}; consequently, structure bending becomes a dominant deformation mode. To capture this behavior, the bending stiffness of the continuous structure is lumped into discrete bending hinges placed along the shared edges between adjacent triangular facets. For example, two adjacent triangular facets with areas $A_1$ and $A_2$ sharing a common edge of length $L$, undergo a relative rotation by an angle $\theta$ upon loading, as illustrated in Figure~\ref{Hinge_description}c.

Each bending hinge element represents the bending response of a structure with thickness $t$ over a representative bending length $c$. Here, $c$ corresponds to the circular arc length formed after bending the rectangular strip associated with the two adjacent facets, as shown in Figure~\ref{Hinge_description}c. It is assumed that the portion of the structure contributing to this bending deformation—namely, the region associated with the two adjacent triangles—is relatively narrow due to a fine mesh discretization. Under this assumption, the curvature within the hinge region is approximately constant and denoted by $k$.

The bending hinge stiffness is derived by equating the bending moment of a continuous elastic structure to that of an equivalent discrete hinge. This derivation incorporates both the geometric characteristics of the mesh and the material properties of the structure. For a slender elastic structure, the linear moment–curvature relation is given by

\vspace{-3mm}
\begin{equation}
    M_{cont} = EIk,
    \label{eq:Bending_moment}
\end{equation}

\noindent where $I$ is the second moment of area and $k$ is the bending curvature. For the considered scenario (shown in Figure \ref{Hinge_description}), substituting $I$ and $k$ into Eq. \ref{eq:Bending_moment} yields

\vspace{-3mm}
\begin{equation}\label{M_cont}
    M_{cont} = E \left(\frac{Lt^3}{24} \right) \left( \frac{\theta}{c} \right),
\end{equation}

\noindent where $L$ is the common edge length, $\theta$ is the dihedral rotation between adjacent facets, and $c$ is the characteristic distance between facet centroids. 
Utilizing the discrete bending hinge formulation, 

\vspace{-3mm}
\begin{equation}\label{M_desc}
    M_{desc} = k_B\theta,
\end{equation}

\noindent where $k_B$ is the bending hinge stiffness. Equating Eqs. \ref{M_cont} and \ref{M_desc}, 

\vspace{-3mm}
\begin{equation}
    k_B = \frac{ELt^3}{24c}.
\end{equation}

To relate the discrete hinge model to continuum plate bending, the pair of adjacent triangular facets sharing an edge of length $L$ is replaced by an equivalent rectangular strip having the same area and thickness. The rectangle is assigned a width of L/2 and the effective strip length $c$ is obtained as $2(A_1+A_2)/L$. After substituting $c$ the final bending stiffness expression

\vspace{-3mm}
\begin{equation}\label{eq:hinge_stiffness}
    k_B = F_B  \frac{E{L}^2t^3}{48(A_1+A_2)}.
\end{equation}

This formulation assumes that the bending behavior of the adjacent triangular facets can be represented by an area-equivalent rectangular strip. The approximation is generally accurate for sufficiently refined meshes, where the triangles provide a good local representation of the underlying surface. For nonuniform or coarse meshes, the approximation remains energy-equivalent but may lose accuracy due to the larger, irregular elements. In the present work, this error is compensated through the calibration factor $F_B$ which is determined through an iterative calibration procedure using high-fidelity Abaqus simulations as the reference solution. A representative set of nearly 15 kirigami structures with different geometries were analyzed. The value of $F_B$ was systematically varied until the discrepancy in the global force-displacement response between the bar and hinge model and the Abaqus simulations was minimized. In practice, the reaction force was used as the primary calibration metric because it exhibited the highest sensitivity to variations in the bending stiffness. The optimal value of $F_B$ was found to be 0.125, and this value is used for all examples presented in this work.

The bending hinge contribution to the global stiffness matrix is assembled as $\mathbf{K}_B = \mathbf{J}_B^T \mathbf{D}_B \mathbf{J}_B$. 

\subsection{Three-node torsional spring stiffness}
The rotational spring element is employed exclusively for in-plane, two-dimensional kirigami analysis. In in-plane kirigami systems—such as structures programmed to deploy into prescribed shapes \cite{choi2019programming} and topologically polarized kirigami lattices \cite{dang2025kirigami}—bar elements that capture only in-plane axial and shear responses are insufficient to represent the full mechanical behavior. To address this limitation, a three-node torsional spring (3NTS) element is introduced between pairs of connected bars to model rotational stiffness. This element resists changes in the relative angle between adjacent bars and implicitly introduces rotational deformation without the need for explicit rotational degrees of freedom. The 3NTS element consists of two elastic axial force members (bars) joined at an intermediate node, which hosts a torsional spring of lumped rotational stiffness $\mathbf{K}_R$, as shown in Figure~\ref{Bar_&_hinge_discretization}(b3). 

The rotational spring formulation adopted here follows the three-node torsional spring framework developed in \cite{patil2026three}. The stiffness matrix of a two-dimensional 3NTS is derived from the spring's constitutive law by evaluating the Jacobian and Hessian of its strain energy. The strain energy of a rotational spring is defined as
\begin{equation}
U(\alpha) = \frac{1}{2} k_R \left( \alpha - \alpha_0 \right)^2 ,
\label{eq:rotational_energy}
\end{equation}
where $k_R$ is the rotational stiffness, $\alpha$ is the relative angle between two connected bars, and $\alpha_0$ is the rest angle in the undeformed configuration. The relative angle $\alpha$ is a function of the global nodal coordinates, allowing the rotational spring to be expressed entirely in terms of translational degrees of freedom.

Linearization of the angle measure yields the following compatibility relation
\begin{equation}
\alpha - \alpha_0 = \mathbf{J}_R \mathbf{u},
\label{eq:rotational_compatibility}
\end{equation}
where $\mathbf{J}_R$ is the Jacobian matrix of the relative angle with respect to the global nodal displacement vector $\mathbf{u}$. The corresponding moment--rotation constitutive relation is given by
\begin{equation}
M = k_R \left( \alpha - \alpha_0 \right),
\label{eq:rotational_constitutive}
\end{equation}
which leads to the following matrix-based stiffness contribution
\begin{equation}
\begin{aligned}
\mathbf{K}_R &= \mathbf{J}_R^{\top} \mathbf{D}_R \mathbf{J}_R, \quad
\mathbf{D}_R = \mathrm{diag}\!\left( k_R \right).
\end{aligned}
\label{eq:rotational_stiffness}
\end{equation}

Details of the full derivation, including higher-order nonlinear terms, are provided in \cite{patil2026three} and are not repeated here for brevity.

\section{Bar and hinge discretization, boundary constraints, and non-linear solver}
\label{sec:discretization_BC}

 \begin{figure*}
    \centering
    \includegraphics[width=0.95\textwidth]{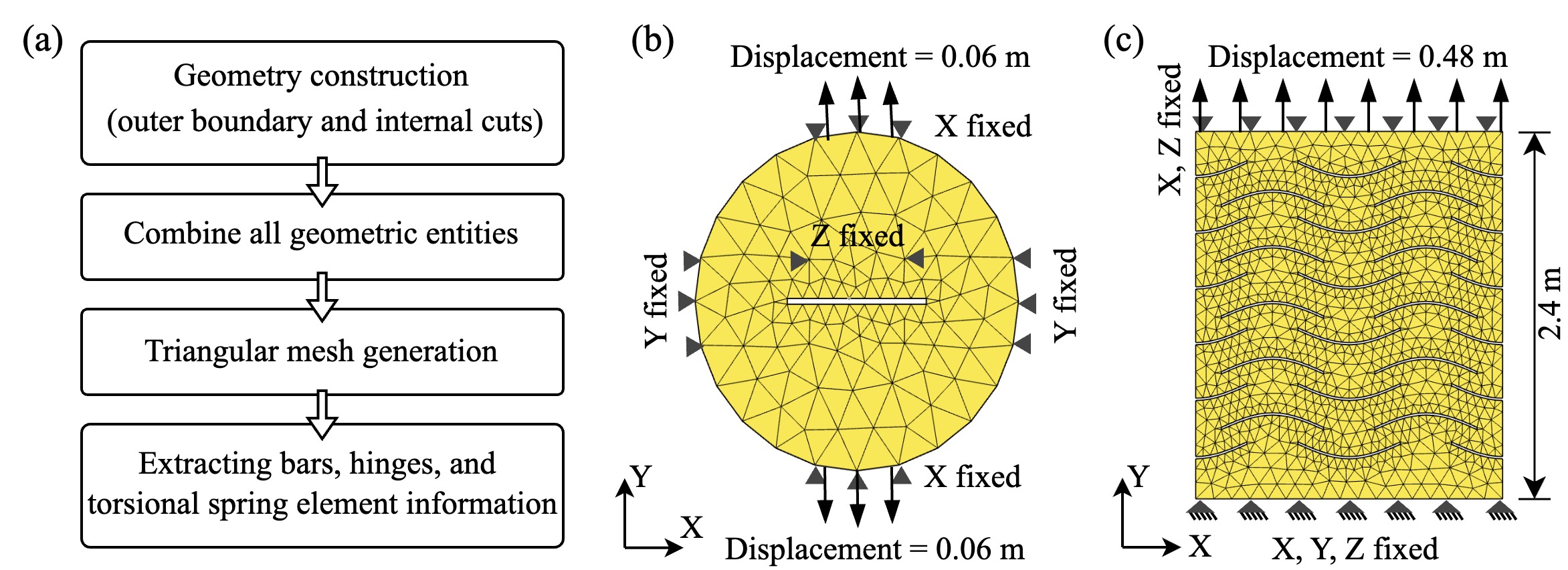}
    \caption{(a) Bar and hinge descretization process. (b, c) Applied boundary conditions for stretching simulations of circular and rectangular kirigami structures.}
    \label{mesh_pipeline_&_BC}
\end{figure*}

The kirigami structure is discretized using a triangular mesh to accommodate bar, hinge, and torsional spring elements. This discretization process is outlined in Figure \ref{mesh_pipeline_&_BC}a. The geometry is first constructed using the MATLAB PDE Toolbox by defining the outer sheet boundary and internal cuts in the decsg (decomposed constructive solid geometry) format. These geometric entities are combined into a single geometry description matrix, in which the cut regions are subtracted from the outer sheet domain using the Boolean operation $'O_1-C_1'$. 
The resulting geometry is then imported into a PDE model and discretized using a triangular FE mesh.

The mesh discretization is controlled solely by the prescribed maximum edge length, $L_{max}$, which determines the size of the triangular facets and, consequently, the length of the bars and bending hinges formed along their common edges. The value of $L_{max}$ is selected to balance surface curvature and computational efficiency.
No additional local mesh refinement is employed near hinges. However, local refinement is performed automatically near cut edges to improve numerical accuracy and the prediction of out-of-plane deformation. Finally, the bar and hinge connectivity is extracted from the triangular mesh for three-dimensional simulations, whereas the bar and torsional spring connectivity is used for in-plane analyses.

Appropriate boundary conditions are applied to the considered physical systems, with careful treatment to eliminate rigid body modes while avoiding over-constraining the model. Figures~\ref{mesh_pipeline_&_BC}b and~c illustrate the boundary conditions used in the simulations discussed in the following section. The cases include a circular structure subjected to stretching in two opposing directions and a rectangular structure fixed at one end and stretched from the other. A reference node or a set of boundary nodes is constrained to suppress global translation, while additional constraints are introduced to prevent rigid body rotation. This approach preserves the intrinsic deformability of the kirigami structure while ensuring a well-conditioned stiffness matrix for static and stability analyses.


The proposed formulation considers geometric nonlinearity arising from large displacements and rotations while assuming linear elastic material behavior. The constitutive response of the bars, bending hinges, and torsional springs is defined using constant elastic properties, and no material nonlinearity is included. This assumption is appropriate for the thin-sheet kirigami structures considered in this work, where the global mechanical response is primarily governed by geometric changes during deformation.
To solve the geometrically nonlinear kirigami system, an incremental–iterative displacement-controlled scheme, implemented within a Newton–Raphson framework and combined with adaptive step sizing, is employed. This approach enables robust tracing of the nonlinear response associated with large deformations and instability-driven behavior commonly observed in kirigami structures. 

Additionally, a small initial geometric perturbation is introduced in the undeformed configuration during preprocessing. Thin-sheet kirigami structures are inherently sensitive to geometric imperfections, which can influence the onset and direction of out-of-plane buckling. Such behavior is well documented in experimental studies and has been highlighted in \cite{chaudhary2023geometric}. Accordingly, the introduction of a controlled and small perturbation facilitates the capture of physically relevant responses. The influence of perturbation type and magnitude on the model response is examined in the following section.

\section{Results and discussion}
This section discusses the impact of certain parameters such as, descretization size and initial perturbation on the mechanics and kinematics of kirigami. Furthermore, the accuracy and efficiency of the bar and hinge model is discussed by comparing the results obtained from commercial software and experimental testing. 

\begin{figure}
    \centering
    \includegraphics[width=0.99\textwidth]{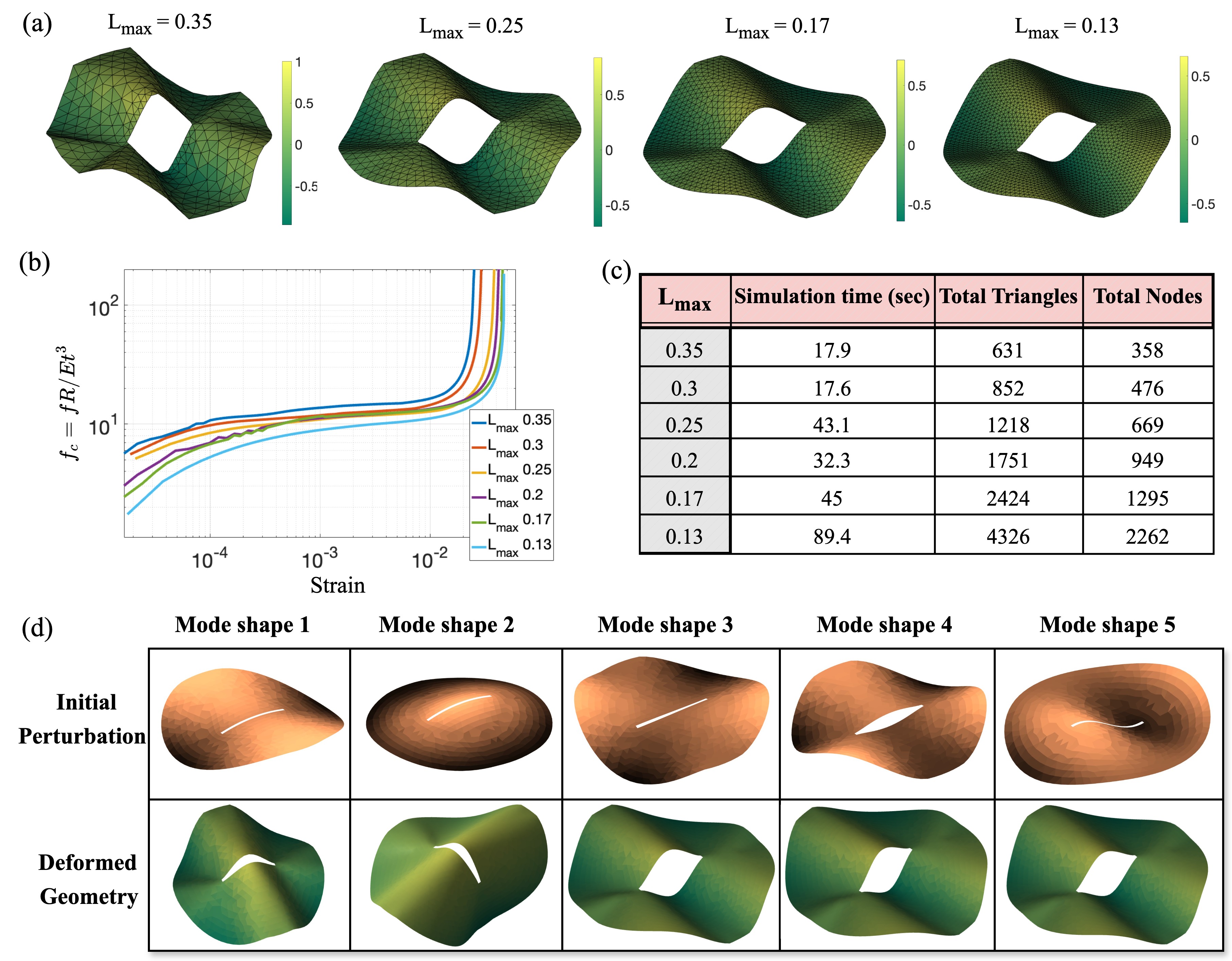}
    \caption{Effect of mesh size and initial perturbation on the deformed shapes and stiffness of kirigami. (a) Mesh size effect on the deformed shape of the kirigami. $L_{max}$ is the mesh control parameter that defines maximum allowable length. (b) Force-displacement plot for six different mesh size. We observe a convergence in deformed shape and stiffness after a certain mesh size. (c) computational time and total elements details for six different mesh sizes. (d) Effect of initial geometry/perturbation on the deformed shape. Initial geometry is only perturbed in z-direction with maximum of $10^{-2}$ unit length. Perturbations are exaggerated for clarity.}
    \label{Effect_Initial_params}
\end{figure}

\subsection{Effect of mesh size and initial conditions}
\label{initialCondition_effect}
The robustness of the model can be evaluated by studying the effects of variations in mesh and initial perturbation. To understand the influence of these parameters on the deformed shapes and stiffness of kirigami, a circular structure of radius 3 meters with a single cut at the center is considered. Further details of the structure and boundary conditions are provided in Figure~\ref{mesh_pipeline_&_BC}b. Six mesh sizes with maximum allowable element lengths $(L_{max})$ ranging from 0.13 to 0.35 are examined, as shown in the table in Figure \ref{Effect_Initial_params}c. The deformed geometries after stretching for four different mesh sizes are presented in Figure \ref{Effect_Initial_params}a, and the corresponding element details are listed in Figure \ref{Effect_Initial_params}c.
As observed, both the coarsest and finest meshes produce similar overall configurations; however, the surface curvature and structural stiffness converge as the mesh is refined. This convergence is most evident in the force-displacement curves (Figure~\ref{Effect_Initial_params}b), where the responses exhibit a similar strain region corresponding to the onset of the sharp stiffness increase. Additionally, the computational time remains relatively low for both coarse and fine meshes, indicating the bar and hinge model's ability to achieve fast simulations even at higher mesh resolutions.

The mechanical response of the model is illustrated by the reaction force–displacement curve in Figure \ref{Effect_Initial_params}b. Here, $f_c$ denotes a dimensionless parameter representing force. The plot is presented on a log–log scale, which may give the appearance of tighter clustering of lines. However, convergence is observed beyond a certain level of discretization, both at the buckling point and at the onset of the sharp increase in force. This behavior is primarily attributed to the improved smoothness of the fold curvature. The plateau in the force–displacement curve indicates the onset of out-of-plane buckling, characterized by large displacements with minimal increases in force.
Overall, due to the complex curvature patterns in kirigami structures, relatively finer meshes are often required to accurately capture each fold and curvature. The bar and hinge model demonstrates sufficient robustness and computational efficiency to capture these geometric intricacies effectively.

To examine the effect of initial perturbations in the bar and hinge model, small geometric perturbations were intentionally introduced into the initial geometry, and their influence on the final deformation under stretching was analyzed. Figure~\ref{Effect_Initial_params}d illustrates the resulting deformed shapes for various initial perturbations. These perturbations correspond to the eigenmodes of the initial geometry and were applied by modifying only the Z-coordinates, with a maximum amplitude of $10^{-2}$ times the circular radius (enlarged in the figure for clarity). The results show that while the model generally reproduces the intended deformed shapes, certain modes (e.g., mode shapes 1 and 2) lead to distinct deformation patterns, demonstrating the model’s sensitivity to initial geometry and its robustness in capturing such perturbation effects.

\subsection{Bar \& hinge and commercial software comparison for cellular kirigami structures}

\renewcommand{\arraystretch}{1.5} 
\begin{table}[]
\begin{tabular}{|
>{\columncolor[HTML]{E6E6E6}}c |cc|cc|c|}
\hline
\cellcolor[HTML]{F8CECC} & \multicolumn{2}{c|}{\cellcolor[HTML]{F8CECC}\textbf{Bar \& Hinge model}} & \multicolumn{2}{c|}{\cellcolor[HTML]{F8CECC}\textbf{Abaqus model}} & \cellcolor[HTML]{F8CECC} \\ \cline{2-5}
\multirow{-2}{*}{\cellcolor[HTML]{F8CECC}\textbf{\begin{tabular}[c]{@{}c@{}}Kirigami\\ Geometry\end{tabular}}} & \multicolumn{1}{c|}{\cellcolor[HTML]{F8CECC}\textbf{Total nodes}} & \cellcolor[HTML]{F8CECC}\textbf{Simulation time (sec)} & \multicolumn{1}{c|}{\cellcolor[HTML]{F8CECC}\textbf{Total nodes}} & \cellcolor[HTML]{F8CECC}\textbf{Simulation time (sec)} & \multirow{-2}{*}{\cellcolor[HTML]{F8CECC}\textbf{\begin{tabular}[c]{@{}c@{}}Open Area \\ (\% difference)\end{tabular}}} \\ \hline
$2\times5$ cells & \multicolumn{1}{c|}{7542} & 89.4 & \multicolumn{1}{c|}{67325} & 1144 & 4 \\ \hline
$5\times10$ cells & \multicolumn{1}{c|}{11581} & 102.1 & \multicolumn{1}{c|}{69453} & 1835 & 4.5 \\ \hline
$5\times15$ cells & \multicolumn{1}{c|}{14068} & 185.9 & \multicolumn{1}{c|}{71929} & 4352 & 3.8 \\ \hline
$10\times10$ cells & \multicolumn{1}{c|}{13704} & 165.9 & \multicolumn{1}{c|}{75899} & 81645 & 6.8 \\ \hline
\end{tabular}
\caption{Quantitative comparison between bar and hinge and shell element Abaqus model for cellular kirigami.}
\label{tab:qualitative-comparison}
\end{table}

To demonstrate the effectiveness and efficiency of the bar and hinge model for the analysis complex kirigami structures, comparisons are made with results obtained from Abaqus FE simulations for various cellular configurations. 
The Abaqus simulations follow the modeling procedure described in \cite{arauz2023evaluation}. 
The component was modeled as an isotropic, homogeneous, linear elastic shell structure with a Young's modulus of 2.5 GPa, a Poisson's ratio of 0.34, and a density of 1420 kg/m$^3$. The structure was modeled using 3-node triangular general purpose shell elements. A nonlinear implicit dynamic analysis was performed to approximate quasi-static loading, with a prescribed displacement of 0.2 m applied linearly over 600 s. Numerical damping was introduced using the Hilber--Hughes--Taylor time integration method with $\alpha=-0.41421$, $\beta=0.5$, and $\gamma=0.91421$. Initial geometric imperfections were introduced through an elastic eigenvalue analysis to trigger the out-of-plane buckling response. Standard convergence tests were performed to ensure that the mesh size was sufficiently small to produce accurate results for all considered design cases.

\begin{figure}
    \centering
    \includegraphics[width=0.98\textwidth]{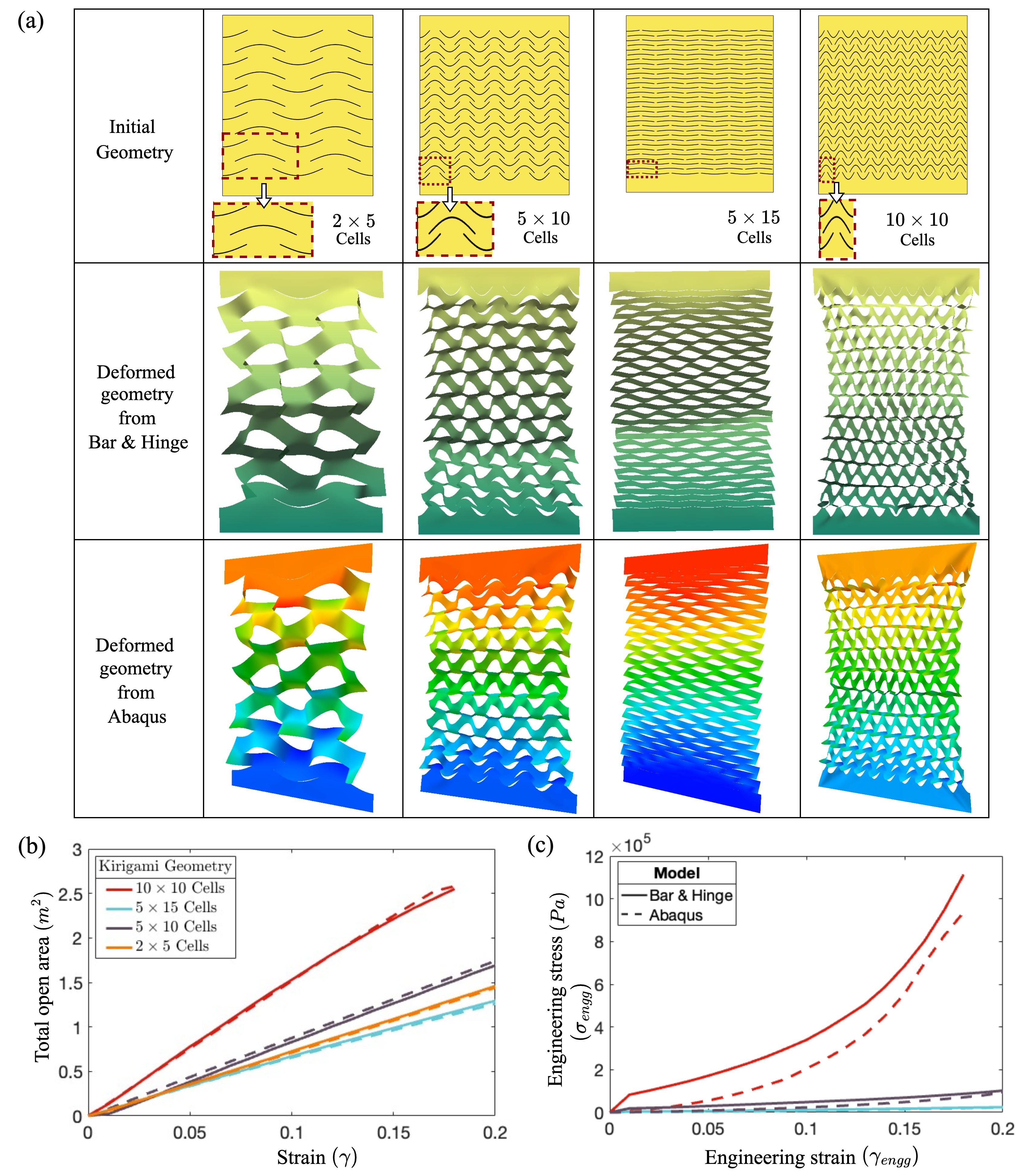}
    \caption{The deformed shapes and stiffness comparison between the bar and hinge model and the shell-element Abaqus model. (a) Initial and deformed geometries of four different cellular structures. Unit cells are indicated by red dashed lines. (b) Total open area versus strain for the structures shown above. (c) Reaction force versus strain for the same structures. Dashed and solid lines represent the Abaqus and bar and hinge models, respectively. Colors represent different number of cells.}
    \label{Cellular_Comparison}
\end{figure}

As shown in Figure~\ref{Cellular_Comparison}a, four distinct cellular kirigami patterns are considered. For example, these patterns have been considered for adaptive building facades \cite{arauz2025comprehensive}, where controlled cut openings regulate the internal temperature of a building. 
The selected designs vary in the number of cells and cut curvature to represent a range of geometric complexities—from few, wide cuts to numerous small ones.
Each structure is stretched by $20\%$ of its length in the direction transverse to the cuts. Further details regarding the geometry and boundary conditions are provided in Section~\ref{sec:discretization_BC}. The deformed shapes from the bar and hinge model and the Abaqus simulation are compared in Figure~\ref{Cellular_Comparison}a, where the color gradient represents deformation in the actuation direction. The bar and hinge model captures the overall deformation well, exhibiting similar global behavior with only minor local variations. Specifically, features such as cut-opening shapes, out-of-plane buckling curvatures, lateral contraction/expansion, and overall deformation closely match the FE results. Small discrepancies appear in a few local out-of-plane buckling and cut-opening directions.

For a quantitative comparison of deformation behavior and mechanical responses, the total cut open area and reaction forces are compared in Figures~\ref{Cellular_Comparison}b and c, respectively. The solid and dashed lines represent the bar and hinge and Abaqus models. The cut opening area is defined as the total area generated by the opening of all cuts upon stretching, evaluated from the front view.
The open area–strain curves (Figure~\ref{Cellular_Comparison}b) show less than a $6\%$ difference between the two models for all considered cut patterns. Detailed open area comparisons and model specifications for each kirigami design are provided in Table~\ref{tab:qualitative-comparison}, which also lists the total number of nodes and computational time for both approaches. The model details and computational times are obtained from the Abaqus report files and MATLAB runs for the respective models. All simulations were performed on a personal computer equipped with an Apple M2 chip and 64 GB of RAM. For all cases, the bar and hinge model achieves over an order-of-magnitude improvement in computational efficiency, primarily due to its reduced number of degrees of freedom; in addition, improved convergence characteristics lead to fewer iterations to reach convergence. In contrast, the Abaqus shell-element model requires a finer mesh and smaller displacement increments to ensure convergence. Due to the larger number of cuts, the $10\times10$ cell kirigami structure is stretched only up to a strain of 0.17 in both Abaqus and bar and hinge models, which is why its curve appears shorter than the others in Figures~\ref{Cellular_Comparison}b and c.

The engineering stress–strain curves (Figure~\ref{Cellular_Comparison}c) demonstrate the ability of the bar and hinge model to capture the stiffness behavior with good agreement to the Abaqus simulations. The engineering stress is calculated using the gross initial cross-sectional area of the specimen, thereby characterizing the effective macroscopic response of the kirigami architecture.
Overall, both models exhibit similar trends, although the bar and hinge model generally predicts slightly higher stiffness. In the more flexible structures (e.g., $5\times15$ and $5\times10$ cells), the agreement is closer, while for stiffer 


\clearpage

\begin{figure}
    \centering
    \includegraphics[width=0.98\textwidth]{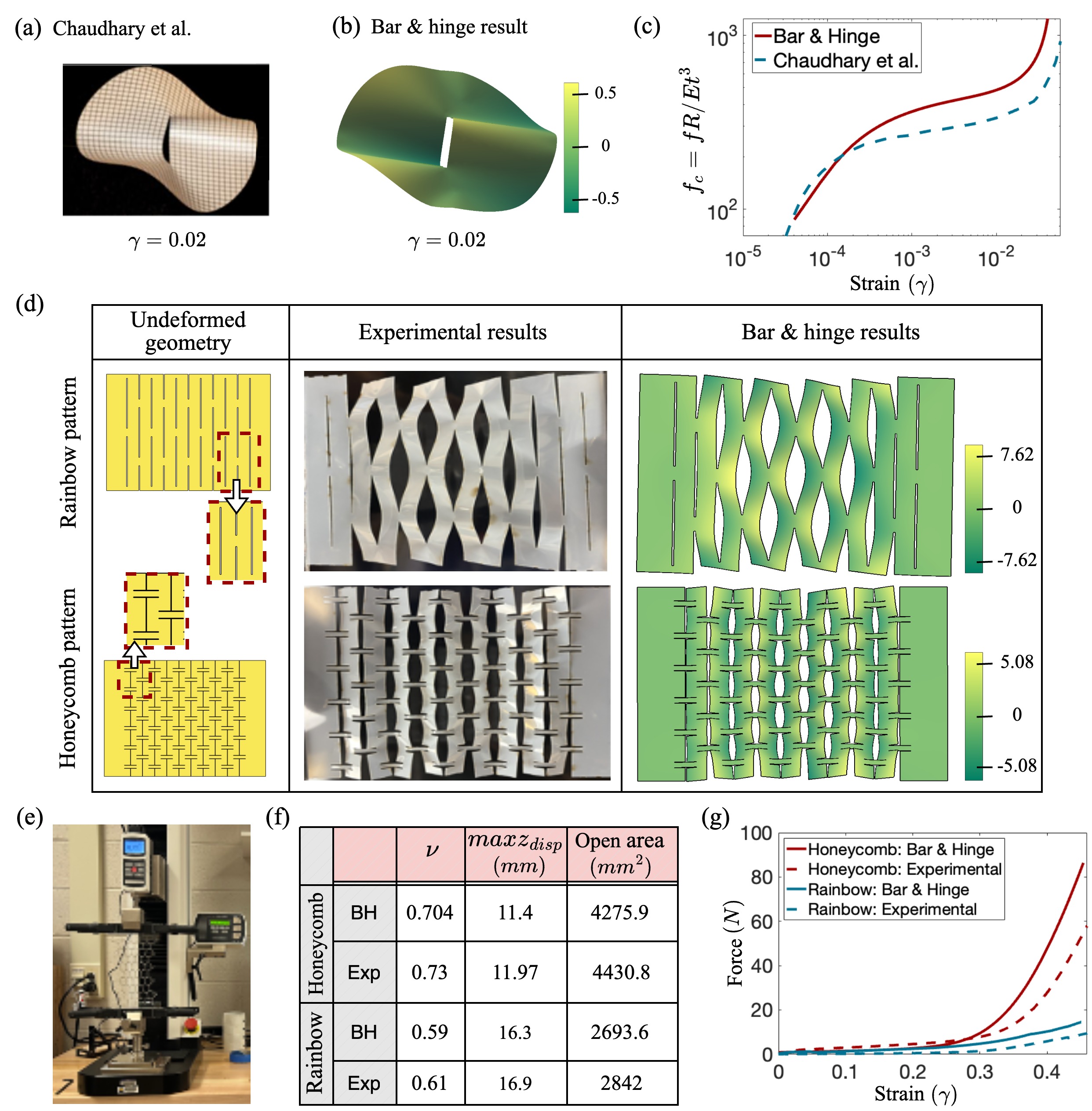}
    \caption{Bar and hinge model comparison with experimental results. (a, b) comparison of deformation behavior for a single-cut circular structure stretched to a strain of 0.02. (c) Numerically obtained force–displacement curve for the cases shown in (a) and (b). The experimental results are adapted from Chaudhary et al. \cite{chaudhary2023geometric}. (d) Comparison with complex cellular kirigami. Rainbow and honeycomb patterns are considered, with their unit cells outlined by dashed red lines. The color bar in the bar and hinge results indicates deformation in the Z-direction (mm). (e) Kirigami testing setup on the Mark-10 UTM. (f) Quantitative comparison of Poisson’s ratio, maximum Z-displacement, and total cut open area. (g) Force–displacement curve up to a strain of 0.5. Dashed and solid lines represent experimental and bar and hinge responses, respectively.}
    \label{Comp_with_Exp}
\end{figure}
\clearpage

\noindent configurations (e.g., $10\times10$ cells), the bar and hinge model shows a somewhat higher stiffness response. 
Additionally, unlike the single-cut kirigami simulations, the presence of multiple cuts in these cellular structures causes localized buckling around the cuts to initiate at small strains, as evident from Figure~\ref{Cellular_Comparison}c. 

\subsection{Comparison with experimentally tested structures}
To assess the accuracy of the bar and hinge model, a comparison with experimentally tested kirigami structures is presented. Figure~\ref{Comp_with_Exp} shows the results for single- and multi-cut configurations. As reported in the literature \cite{chaudhary2023geometric,dang2025kirigami}, single-cut kirigami exhibits more complex deformation with highly nonuniform out-of-plane curvature arising from buckling. Here, the same circular structure with a single cut is considered as shown in Section~\ref{initialCondition_effect}. Details of the geometry and boundary conditions are provided in Section~\ref{sec:discretization_BC}. Experimental data from Chaudhary et al. \cite{chaudhary2023geometric} are used for validation.

The bar and hinge model accurately captures the global deformation response—including overall curvature and cut opening—with only minor local deviations, as shown in Figures~\ref{Comp_with_Exp}a and b for a strain of 0.02. The corresponding force–strain curves in Figure~\ref{Comp_with_Exp}c illustrate similar trends before and after buckling, though the bar and hinge model predicts a slightly stiffer response than observed experimentally. The force–strain response exhibits three distinct deformation regimes: (i) an initial linear regime dominated by in-plane stretching, (ii) an out-of-plane buckling regime associated with bending-dominated deformation, and (iii) a post-buckling regime characterized by a sharp increase in force. The third regime reflects structural stiffening as deformation transitions from bending-dominated to stretching-dominated behavior with increasing strain.

For multi-cut configurations, cellular kirigami structures based on rainbow and honeycomb patterns were used, as shown in Figure~\ref{Comp_with_Exp}d. In these designs, straight cuts were arranged in parallel and perpendicular orientations. The samples were fabricated from Mylar sheets using a laser-cutting machine. Experimental tests were conducted on a Mark-10 Universal Testing Machine (UTM), where the top and bottom ends of the sheet were uniformly clamped between two panels and stretched upward, as illustrated in Figure~\ref{Comp_with_Exp}e. To prevent slippage during large deformations, the panels were further secured with clamps and pins. Each specimen was stretched up to a strain of 0.5 (relative to the initial length), which corresponded to the maximum achievable strain before failure.
The bar and hinge model exhibited deformation behavior closely matching the experimental results (Figure~\ref{Comp_with_Exp}d), particularly in terms of cut-opening shapes, lateral deformation, fold curvature, and out-of-plane displacement.

For quantitative comparison of the multi-cut configurations, several parameters from the deformed geometry and the force–strain response are correlated to evaluate the deformation behavior and stiffness accuracy of the bar and hinge model. The deformed geometries from experimental runs were captured using point-cloud data and processed in MATLAB. From these data, the Poisson’s ratio, maximum out-of-plane (Z) displacement, and total open area were computed and compared with the corresponding results from the bar and hinge simulations, as shown in Figure~\ref{Comp_with_Exp}f. The Poisson’s ratio reflects the lateral contraction upon stretching. The bar and hinge model showed less than 4\%, 4\%, and 5\% differences in Poisson’s ratio, maximum Z-displacement, and total open area, respectively.
The stiffness response in Figure~\ref{Comp_with_Exp}g further demonstrates that the bar and hinge model closely follows the experimental trend. It accurately captures the low-strain region, though it becomes slightly stiffer at higher strains. Overall, the bar and hinge model effectively captures both the global deformation and the intricate local configurations.

\subsection{Bar and hinge implementation for non-periodic and irregular cut patterns}
\begin{figure}
    \centering
    \includegraphics[width=0.99\textwidth]{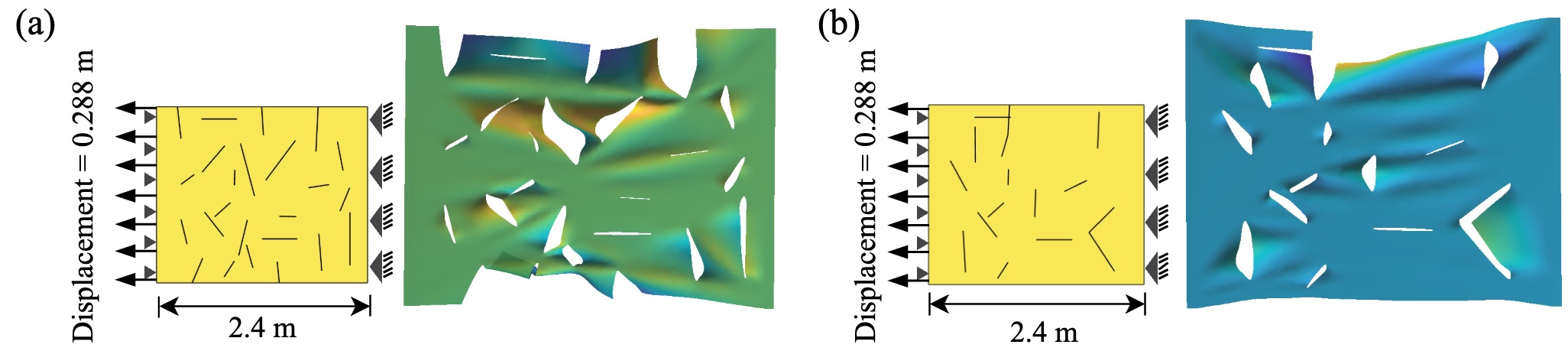}
    \caption{ Bar and hinge simulations of non-periodic cut patterns: (a) non-intersecting cuts and (b) intersecting cuts.}
    \label{Irregular_cuts}
\end{figure}

To demonstrate the broader applicability of the proposed method, its implementation is showcased for non-periodic and irregular cut patterns. The deformation behavior of two representative irregular cut patterns is presented in Figure~\ref{Irregular_cuts}. In both cases, the structure is fixed along one edge and subjected to 12\% tensile strain along the opposite edge. The out-of-plane displacement of the deformed kirigami structures is represented by the color gradients. These examples demonstrate the capability of the proposed bar and hinge formulation to simulate kirigami structures beyond the periodic, regular, and semi-regular cut patterns considered in the preceding examples.

\section{Application problems}
This section highlights the benefits and effectiveness of the bar and hinge model through four representative and widely studied applications of kirigami-inspired structures. These examples encompass parametric studies, complex kirigami simulations, and kirigami-inspired metamaterial analyses, all of which are challenging to implement efficiently using conventional FE software.

\subsection{High-throughput sweeps for property space exploration}
\begin{figure}
    \centering
    \includegraphics[width=1\textwidth]{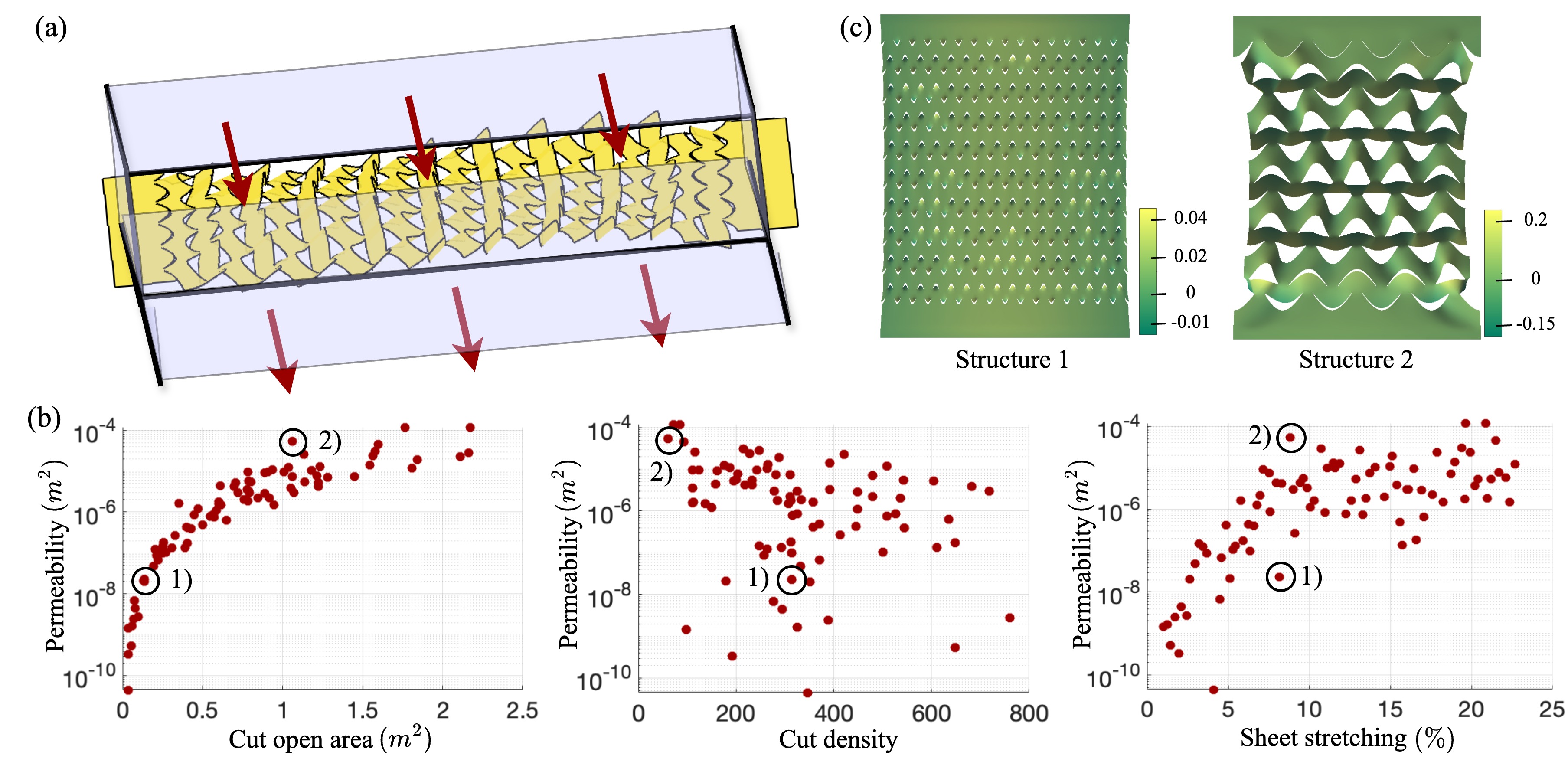}
    \caption{High-throughput parametric study for permeability of kirigami structures. (a) The permeability problem where liquid is flowing from the top to bottom of the enclosed chamber as indicated by red arrows. (b) Permeability versus different geometric parameters. Designs circled with 1 and 2 demonstrate two different structures with distinct geometries and permeabilities. (c) The geometry of these two structures, where the magnitude of out-of-plane deformation is shown with the  color bar.}
    \label{Permeability}
\end{figure}

Any kirigami study that involves simulations over a large number of structures, such as a large parametric study, is difficult and inefficient to simulate using standard FE software. For such scenarios, simplified models work best, as they can provide approximate solutions rapidly, and specific scenarios could be re-evaluated with more advanced FE models as needed. To demonstrate this capability, we implement the bar and hinge model for a large dataset generation problem considering the permeability of a kirigami-inspired structure. Here, a liquid flows from the top to the bottom of a container and through the kirigami structure, as shown in Figure~\ref{Permeability}a. By varying the cut geometry, a wide range of permeability values can be achieved.

The permeability value of the kirigami structure is obtained by connecting a microscopic description of flow through each opened cut (Poiseuille/“cubic-law” behavior for a slit) with the macroscopic Darcy description for the entire structure. Microscopically, each opened cut behaves like a parallel-plate channel whose volumetric conductance scales with the cube of the aperture \cite{neuzil1981flow,rastiello2014real}. Macroscopically, we describe the structure by an intrinsic permeability $k$ that, when combined with the applied pressure gradient, gives the structure-average flux via Darcy’s law \cite{lage1998flow}. Equating the summed slit conductance from Poiseuille behavior to the Darcy flux gives an expression for the structure permeability in terms of the slit geometries \cite{zimmerman1996hydraulic}.

A key modeling choice is the identification of the microscopic channel length (the distance over which a local slit experiences a pressure drop) with the macroscopic Darcy length (the distance used in the Darcy pressure gradient). In this study, both the channel length and the Darcy length correspond to the structure thickness. Since these lengths are equal, they cancel algebraically.
Additionally, several assumptions are adopted: a uniform pressure drop across all cuts, no leakage, laminar flow, and a cut geometry approximated by a near-elliptical shape. The formulas below present the volumetric flow rates predicted by the Darcy ($Q_{\text{Darcy}}$) and Poiseuille ($Q_{\text{Poiseuille}}$) flow models, and the kirigami structure permeability ($k_{\text{sheet}}$) is obtained by equating the two flow rates.
            
\begin{equation} \label{eq:darcy}
Q_{Darcy} = -\frac{k_{sheet} A_{sheet}}{\mu L_{\text{Darcy}}} \Delta P,
\end{equation}

\begin{equation} \label{eq:poiseuille}
Q_{Poiseuille} = - \frac{w^3}{12\mu}  \cdot \frac{l_{\text{cut}}}{L_{\text{channel}}} \cdot \Delta P,
\end{equation}

\begin{equation} \label{eq:final}
\begin{aligned}
k_{\text{sheet}} &= \frac{1}{A_{\text{sheet}}} 
\left( \sum_{i=1}^{N} \frac{w_i^{3}\,l_{cut,i}}{12} \right), & 
w_i = 2\frac{A_{cut,i}}{P_{cut,i}}.
\end{aligned}
\end{equation}

Here, $\mu$ is the dynamic viscosity of the fluid, and $A_{\text{sheet}}$ is the total area of the structure through which flow occurs. $L_{\text{Darcy}}$ and $L_{\text{channel}}$ represent the characteristic flow lengths in the Darcy and Poiseuille formulations, respectively (typically corresponding to the structure thickness). $\Delta P$ is the pressure difference applied across the structure. The quantities $w_i$ and $l_{\text{cut},i}$ denote the effective slit width and the in-plane length of the $i^{\text{th}}$ cut after deformation, respectively. $A_{\text{cut},i}$ and $P_{\text{cut},i}$ are the area and perimeter of the $i^{\text{th}}$ cut used to estimate the effective width, and $N$ is the total number of cuts in the kirigami structure.

This permeability formulation enables direct computation of the structure's permeability from the deformed kirigami geometry obtained using the bar and hinge model.
Key design variables—including the number of cells in the transverse and longitudinal directions, cut spacing in both directions, and percent stretching—were randomly sampled over a wide design space using Latin hypercube sampling \cite{shields2016generalization}, and permeability was computed for each configuration. In total, 75 distinct kirigami designs were analyzed. The effects of cut open area, cut density, and structure stretching on permeability are shown in Figure~\ref{Permeability}. The results reveal a strong correlation between permeability and cut open area, following an approximately cubic relationship, which indicates a nonlinear increase in flow capacity with increasing opening of the kirigami cuts. Notably, structures with identical stretching (Figure~\ref{Permeability}c) can exhibit a wide range of permeability values, governed primarily by total open area rather than the number of cuts. Taken together, these results demonstrate the capability of the bar and hinge model for efficient large-scale dataset generation and property-space exploration. The entire parametric study, including the simulations and permeability computations, took nearly five hours on an Apple M2 Max with 64 GB of RAM. Increasing the size of the parametric study would increase the runtime approximately linearly.

\subsection{Simulation of a kirigami skinned crawler}

Kirigami-skinned crawlers exploit the distinctive mechanical response of patterned sheets that undergo out-of-plane buckling to generate directional and locomotion. In this study, a thin kirigami structure is folded into a three-sided crawler body, with the bottom face serving as the primary contact interface with the substrate. Upon tensile actuation of the structure, the cut patterns on the lower surface open and deform, forming asymmetric protrusions (“fins”) that interact with the ground in a direction-dependent manner. This geometric and mechanical asymmetry results in anisotropic frictional behavior, which is essential for enabling forward locomotion in soft, lightweight crawler systems.
Several studies \cite{seyidouglu2025inflatable,khan2025bioinspired,tirado2025multimodal,rafsanjani2018kirigami} have demonstrated multi-directional locomotion in kirigami-based crawler designs. However, most investigations have been limited to experimental demonstrations, with relatively few computational studies due to the challenges associated with modeling thin-sheet mechanics using conventional FE methods. This example illustrates the implementation of the bar and hinge framework as an efficient approximation of deformation behavior and frictional response in such complex systems.

To quantify the frictional performance, an empirical relation proposed in \cite{khan2025bioinspired} is adopted. Using this approach, the frictional response is estimated based on the geometric parameters of the deformed crawler configurations. Three distinct kirigami geometries—triangular, circular, and trapezoidal cut patterns—are considered in this study. Each pattern alters the local deformation mechanisms, surface engagement, and the density of the pop-out features, thereby influencing the effective frictional behavior. Accordingly, the coefficient of friction for each design is expressed as

\begin{equation} \label{eq:poiseuille}
\mu_f = \mu_0 + K_f \cdot h \cdot \rho \cdot S_f,
\end{equation}

\noindent where $\mu_0$ is the intrinsic base friction coefficient, $K_f$ is an empirical constant capturing the surface engagement characteristics, $h$ is the pop-out height induced by buckling, $\rho$ is the density of cuts, and $S_f$ represents the surface factor corresponding to the contact area ratio. This formulation captures the combined influence of geometric design and mechanical deformation on frictional behavior.


 \begin{figure*}
    \centering
    \includegraphics[width=0.98\textwidth]{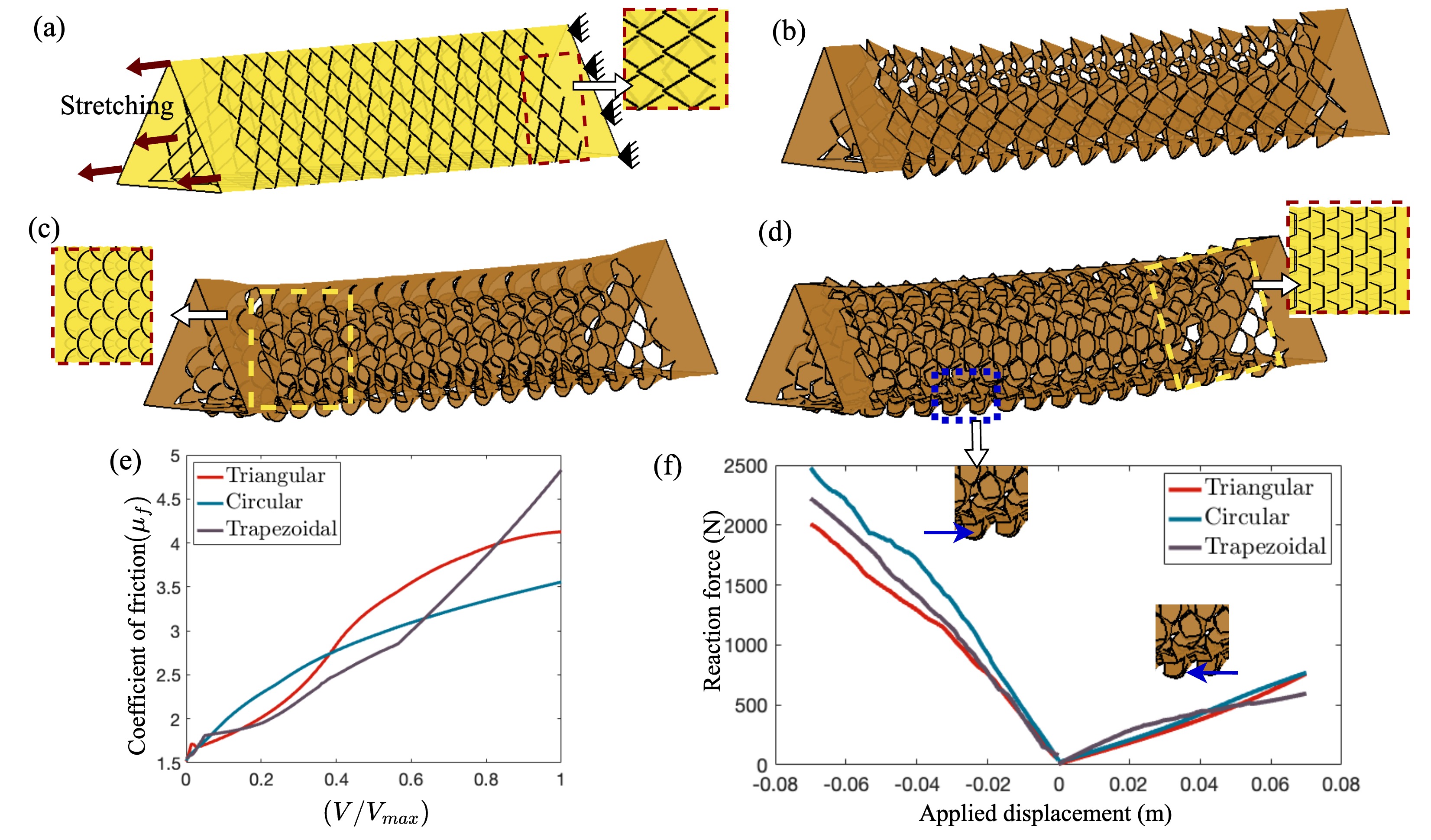}
    \caption{Kirigami skinned crawler simulation. (a) Undeformed geometry with the triangular cut. For the simulation it is fixed on the right side and stretched from the left side. (b) Deformed geometry for triangular cut. (c) Deformed geometry for the circular and (d) trapezoidal cut patterns. (e) Coefficient of friction versus stretching for the three geometries. (f) Reaction force versus applied displacement at a fin. One of the fins is displaced in both the right and left directions and the recorded reaction force shows anisotropic stiffness. The loading region of the fin is illustrated in the plot for the trapezoidal cut pattern.}
    \label{Kirigami_Crawler}
\end{figure*}

The crawler simulation was conducted by fixing the right edge of the folded kirigami structure and applying displacement from the left edge to induce stretching. The resulting out-of-plane deformations for each cut pattern are shown in Figure~\ref{Kirigami_Crawler} (a–d). The triangular cut configuration (Figure~\ref{Kirigami_Crawler}a–b) exhibited the most pronounced directional friction behavior, while the circular and trapezoidal cut patterns (Figure~\ref{Kirigami_Crawler}c–d) demonstrated moderate responses with smoother surface engagement. The variation of the effective friction coefficient with applied strain is shown in Figure~\ref{Kirigami_Crawler}e, highlighting how the deformation-dependent contact mechanics govern the locomotion performance.

To further evaluate the system anisotropy, a directional force response test was performed, where one of the deformed fins was displaced alternately to the right and left, and the corresponding reaction forces were recorded. The resulting force–displacement relationship (Figure~\ref{Kirigami_Crawler}f) confirms the asymmetric resistance that underpins the crawler’s forward movement mechanism. Overall, the results highlight the potential of the bar and hinge modeling framework, which enables efficient approximation of deformation behavior and frictional performance in such complex systems.

\subsection{In-plane kirigami simulation for nonreciprocal system}
 \begin{figure*}
    \centering
    \includegraphics[width=0.98\textwidth]{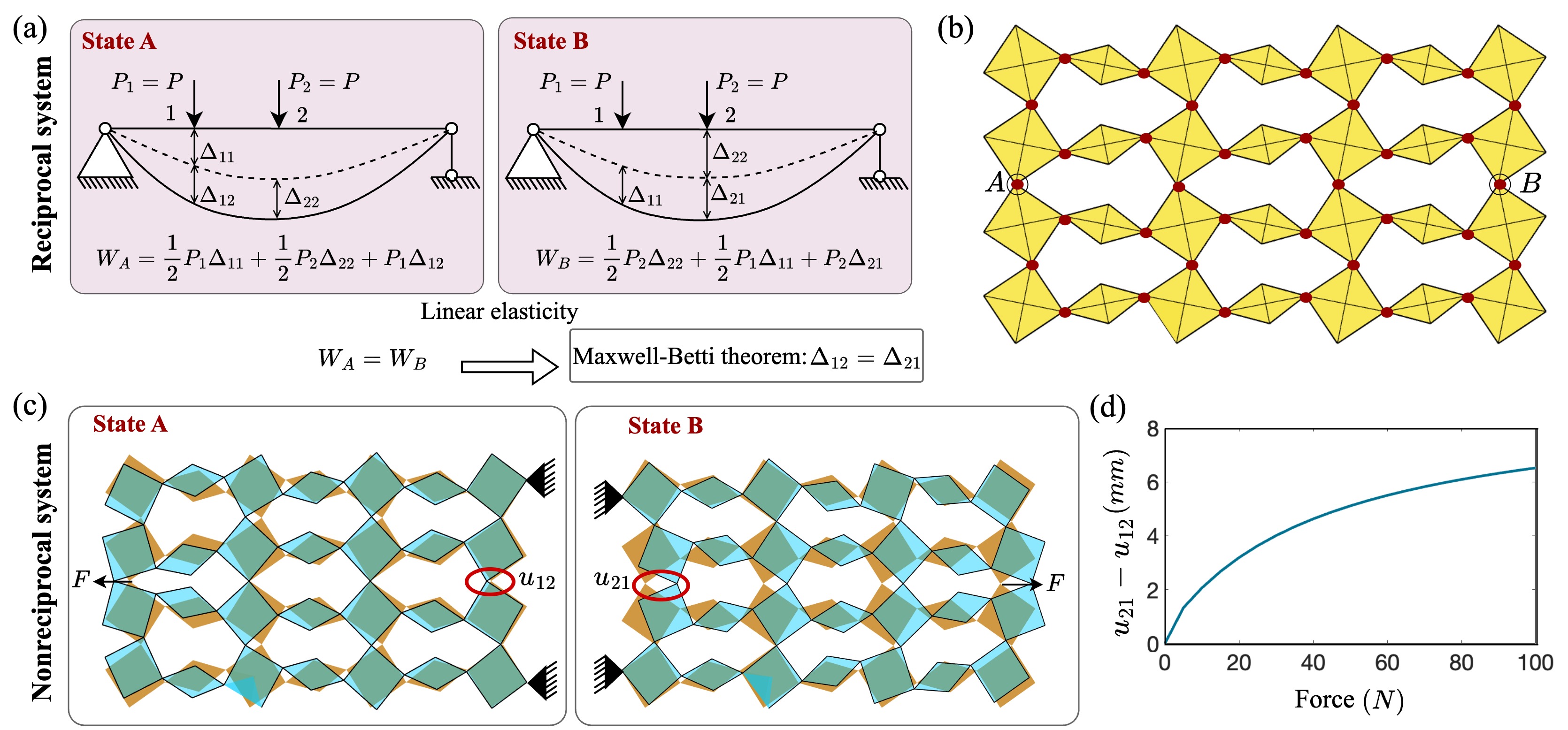}
    \caption{In-plane kirigami simulation using the bar and hinge approach demonstrating static nonreciprocity.
(a) Concept of reciprocity illustrated using a horizontal beam subjected to vertical point loads. Two loading sequences are considered, demonstrating equal deformation in both loading states.
(b) Nonreciprocal system based on a square–rhombus kirigami arrangement with an asymmetric angle $\theta=\pi/16$ \cite{coulais2017static}. Black edges denote bars, and red dots indicate torsional spring elements. Points A and B indicate the loading locations in state A and state B, respectively.
(c) Deformation comparison between state A and state B for the nonreciprocal system. $F$ indicates the point and direction of loading, black triangles represent fixed nodes, and $u_{12}$ and $u_{21}$ denote the resulting deformations in state A and state B, respectively. The brown shading shows the undeformed configuration, while the blue quadrilaterals show the deformed configuration. (d) Displacement difference between state A and state B as a function of the applied load.} 
    \label{Nonreciprocity_figure}
\end{figure*}

Static nonreciprocity refers to the property of a mechanical system in which applying the same static force from opposite sides produces different output displacements, thereby violating the classical Maxwell–Betti reciprocity theorem due to the combined effects of nonlinearity and structural asymmetry \cite{dang2025kirigami}. Coulais et. al., \cite{coulais2017static} proposed a nonreciprocal metamaterial based on a square–rhombus kirigami arrangement with an asymmetric angle $\theta=\pi/16$, as shown in Figure \ref{Nonreciprocity_figure}b.

Static reciprocity can be explained using the Maxwell--Betti theorem, which states that in linear elastic materials under infinitesimal strain, the displacement at one point due to a force applied at another equals the displacement at the second point due to the same force at the first \cite{maxwell1864calculation,charlton1960historical}. As shown in Figure~\ref{Nonreciprocity_figure}a, consider a horizontal beam subjected to vertical forces $P_1$ and $P_2$ at points 1 and 2, respectively, and let $\Delta_{ij}$ denote the displacement at point $i$ due to $P_j$. Two loading sequences are considered: state A, in which $P_1$ is applied before $P_2$, and state B, in which $P_2$ is applied before $P_1$. Under linear elasticity, the total external work is independent of the loading order; thus, $W_A=W_B$ and for $P_1=P_2=P$, reciprocity follows as $\Delta_{12}=\Delta_{21}$.

On the other hand, a square–rhombus kirigami arrangement with asymmetric left and right edges, characterized by an asymmetry angle $\theta=\pi/16$ exhibits static nonreciprocity \cite{coulais2017static}. When identical horizontal forces $F_A=F_B=F$ are applied at the midpoints of the left and right edges (point A and B from Figure \ref{Nonreciprocity_figure}b), the resulting output displacements differ, indicating directional dependence of static transmissibility.  The displacement difference $(u_{21}-u_{12})$ can be seen in Figure~\ref{Nonreciprocity_figure}c (highlighted with red ellipse) and also illustrated in the plot from Figure~\ref{Nonreciprocity_figure}d. Notably, substantial nonreciprocity is observed even for small input forces, $F<10\,\mathrm{N}$. The system is modeled using a Mylar sheet with Young’s modulus E=3.5GPa, and each rhombus has a side length of 6.75mm.  

Here, the bar and hinge model serves as a computationally efficient alternative, and its two-dimensional formulation is employed to obtain the deformed shapes and mechanical response of such structures. The model enables a significantly coarser discretization, as shown in Figure~\ref{Nonreciprocity_figure}(b), while still accurately capturing the deformation and mechanical response, thereby providing a lightweight and computationally efficient framework. This example demonstrates the effectiveness of the bar and hinge formulation in simulating highly nonlinear in-plane kirigami systems.

\subsection{Simulations of in-plane transformable kirigami-inspired metamaterials}

 \begin{figure*}
    \centering
    \includegraphics[width=0.98\textwidth]{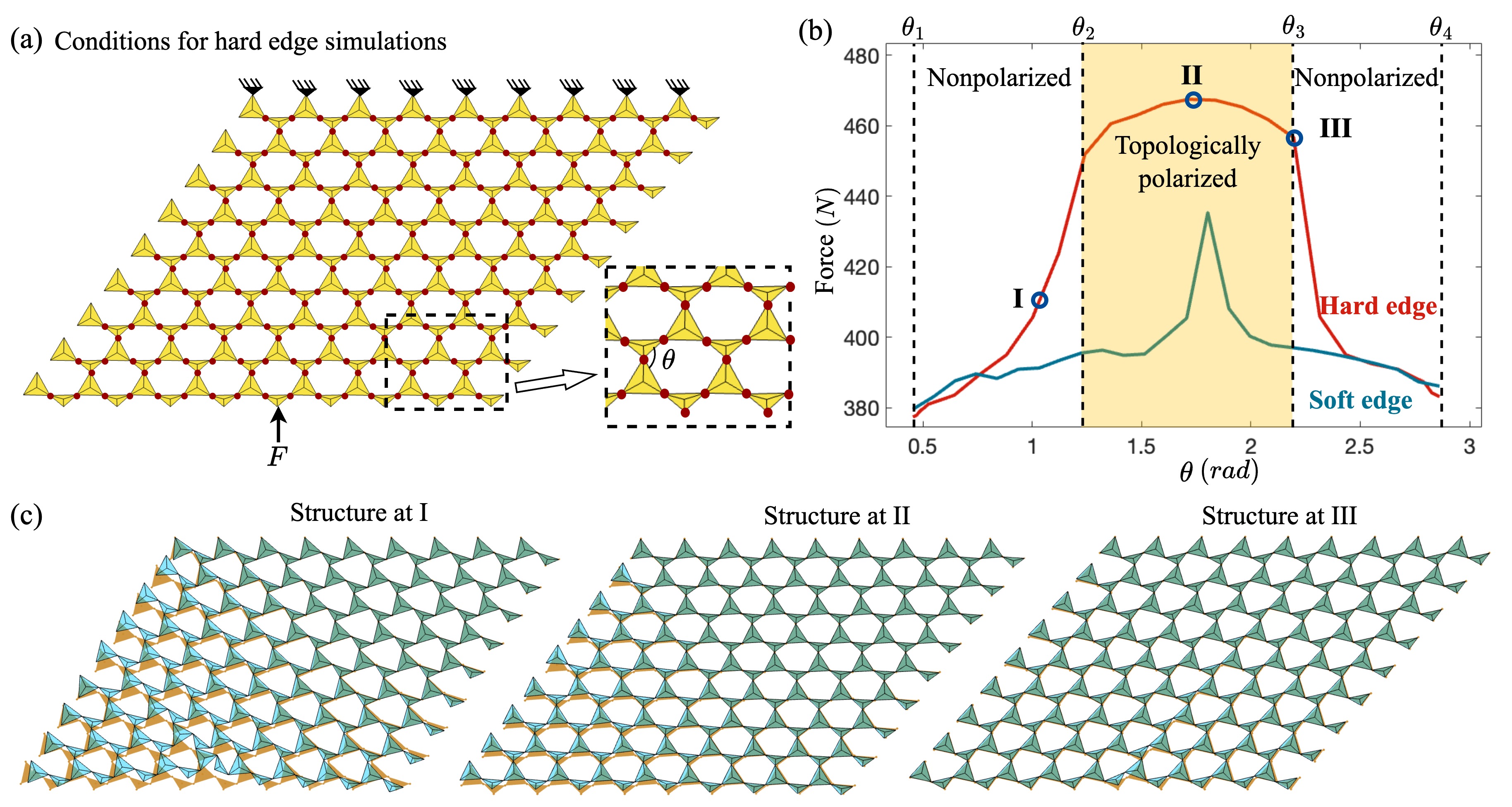}
    \caption{In-plane transformable kirigami-inspired metamaterial simulations using a bar and hinge model. (a) Undeformed configuration and boundary conditions for the hard edge simulation at $\theta=1.92$ radians. Black lines denote bars and red dots indicate the torsional spring elements. The angle $\theta$ is defined between the equilateral and scalene triangles, as shown in the enlarged view. 
    (b) Net reaction force for each twist angle configuration for  both edge conditions. The plot is divided into polarized and nonpolarized regions based on stiffness differences.
    (c) Three representative configurations corresponding to cases I, II, and III. Brown and blue shading represent the undeformed and deformed configurations, respectively.} 
    \label{Top_polarization}
\end{figure*}

This application demonstrates another utility of the bar and hinge model, namely its use for geometric parametric studies of large-scale in-plane cellular kirigami structures. A well-known topologically polarizable kirigami-based metamaterial is reconfigured and simulated. Topological polarization is a structural characteristic whereby pushing the floppy modes from one edge induces additional floppy modes at another edge, and while pushing from the reverse direction does not produce floppy modes \cite{xiu2022topological}. Reconfiguration of the cellular architecture can induce the appearance or disappearance of this polarization, enabling a transformable topological metamaterial \cite{rocklin2017transformable}.

This polarizable structure consists of a cellular arrangement of equilateral and scalene triangles, as shown in Figure~\ref{Top_polarization}a. Each triangle is discretized using six bars and three torsional spring elements. The 3NTS elements are defined following the same manner as in the previous application, where each element is defined by three nodes and an intermediate node, indicated by a red dot. Each configuration is simulated twice and named as either a hard edge or soft edge case depending on its stiffness response. For the hard edge simulations, the top edge is fixed and loading is applied from the bottom, as indicated in Figure~\ref{Top_polarization}a. The boundary conditions are reversed for the soft edge simulations.

The structure is reconfigured by sweeping the twist angle $\theta$ for both soft edge and hard edge simulations, and the corresponding reaction forces are measured to quantify stiffness variations. The twist angle is defined as the angle between the equilateral and scalene triangles, as indicated in Figure~\ref{Top_polarization}a. Figure~\ref{Top_polarization}b shows the net reaction force for each twist angle configuration for both edge conditions. This uniform variation in $\theta$ transforms the structure between polarized states ($\theta_2<\theta<\theta_3$) and nonpolarized states ($\theta<\theta_2$ or $\theta>\theta_3$). As evident in Figure~\ref{Top_polarization}b, the hard edge configuration exhibits a significantly stiffer response in the polarized regime. To illustrate the deformation characteristics in polarized and nonpolarized states, three representative hard edge configurations at $\theta = 1.08$, $1.75$, and $2.2$ radians are shown in Figure~\ref{Top_polarization}c. Each simulation took approximately 20 seconds of computational time. Overall, this study demonstrates that the bar and hinge model captures nonlinear behavior with high computational efficiency and sufficient accuracy, providing a computationally efficient alternative to conventional FE methods for large-scale kirigami metamaterials requiring repeated simulations.

The four application examples collectively demonstrate the versatility of the proposed bar and hinge framework for efficiently analyzing diverse and complex kirigami systems. The current formulation is primarily intended to predict global deformation and reaction forces, and the presented examples therefore focus on deformation behavior and computational efficiency. Nevertheless, the proposed framework provides a foundation for incorporating additional mechanical response metrics, such as energy absorption, stress concentration, or stability characteristics, which will be explored in future developments.

\section{Summary and discussion}
This section summarizes the advantages and limitations of the bar and hinge model for kirigami simulation in comparison with conventional kirigami analysis methods.

\subsection{Advantages of bar and hinge model}
\begin{itemize}[noitemsep, topsep=0pt,label=\footnotesize$\bullet$]
    \item Generalizability: The simplified bar and hinge discretization can be readily applied to a broad class of two- and three-dimensional kirigami-inspired structures. In contrast, theoretical models for approximating kirigami behaviors are derived for specific designs and analytical scenarios. 
    \item Data accessibility: The bar and hinge model provides direct access to stiffness matrices, internal forces, displacements, and other geometrical or mechanical quantities during analysis. Extracting such information from conventional FE software is often nontrivial.
    \item Usage versatility: Owing to its speed and simplicity, the model is well-suited for parametric geometric studies, optimization of cellular kirigami, large-displacement simulations, analysis of large-scale structures, and rapid exploration of eigenvalues and mode shapes.
    \item Computational efficiency: The model employs significantly fewer nodes than a fully discretized FE approach, resulting in a simple and computationally efficient framework.
    \item The model avoids explicit rotational degrees of freedom by recasting resisting moments into equivalent nodal forces consistent with global equilibrium, thereby maintaining a compact stiffness matrix and providing a more simple and understandable formulation.
    \item Ease of use: The formulation is straightforward to understand, modify, and implement, making it accessible to the growing community of kirigami researchers, students, and engineering practitioners.
\end{itemize}

\subsection{Limitations of bar and hinge model}
\begin{itemize}[noitemsep, topsep=0pt,label=\footnotesize$\bullet$]

    \item The model cannot accurately capture localized effects, such as stress concentrations near vertices, and edge effect. These phenomena primarily affect local stress distributions and failure initiation, while having a limited influence on the global force--displacement response and overall deformed configuration that are the primary focus of the present work. To accommodate localized stresses, we can use forces and area of individual bars and can evaluate stress concentration near vertices. 
    However, this will be an extension of this work as it involves extensive study on accuracy improvement and reliability. 

    \item The formulation is currently only available as a MATLAB-based downloadable code \href{https://github.com/Rajkhawale/KirigamiAnalysis-ReducedOrderModel}{[GitHub repository]}. Currently, there is no pre-packaged, user-friendly version of bar and hinge models, which limits accessibility for widespread adoption. 
    \item Shear stiffness is generally overestimated relative to stretching and bending deformations due to the simplified kinematic representation. This slight overestimation of shear stiffness arises from the discrete bar representation, where shear is approximated through axial deformation of the bar network. Although the calibrated factor $F_s$ improves the overall in-plane response, a single global factor cannot completely eliminate shear-related approximation errors. A mode- or mesh-dependent calibration of the bar stiffness could further improve the shear response in future work.
    \item The model requires discretization of the surface, so there are mesh size effects and convergence behaviors similar to conventional FE models.
    \item Bar and hinge models have not been substantially explored for capturing damage in origami, so future work would be needed if these models are to be used to estimate fracture or yielding of the material.
\end{itemize}

\section{Conclusions and future work}


This work presented a generalized and computationally efficient reduced-order model based on the bar and hinge approach for the analysis of kirigami structures. The proposed formulation accurately captures both deformation and internal forces of the system through truss bars, bending hinges, and torsional spring elements. Comparisons with experiments and finite element simulations demonstrated less than $5\%$ error in deformation predictions and less than $10\%$ error in stiffness, while providing nearly an order of magnitude reduction in computational cost. Finally, the advantages and applicability of this model are illustrated through a parametric study for property space exploration, complex analyses such as kirigami-skinned crawlers and simulations for in-plane kirigami-inspired metamaterials.

The proposed bar and hinge kirigami simulation framework establishes a strong foundation for future research and development. Given its computational efficiency, generality, and ability to capture global deformation behavior accurately, the model can be extended in several promising directions. Future work could focus on incorporating material nonlinearity, plastic deformation, and rate-dependent effects to improve prediction fidelity under diverse loading conditions. Integration with optimization and inverse-design algorithms could enable automated exploration of cut-pattern configurations for target mechanical or morphing responses. Coupling the framework with multi-physics modules—such as thermal, electrical, or fluidic interactions—would further broaden its applicability to emerging kirigami-based multifunctional systems. Finally, combining this approach with data-driven techniques, such as machine learning–based calibration or surrogate modeling, could further enhance its predictive accuracy while maintaining low computational cost.

\section*{Availability of the Code Package}
The source code and example files associated with this work are publicly available at:
\href{https://github.com/Rajkhawale/KirigamiAnalysis-ReducedOrderModel}{[GitHub repository]}.

\section*{Acknowledgment}
This work was supported by the U.S. National Science Foundation under the Engineering for Civil Infrastructure program through Award No. 2329760.

\section*{Appendix}

\subsection*{Nomenclature}

\renewcommand{\arraystretch}{0.88}
\setlength{\tabcolsep}{4pt}

\begin{longtable}{p{2.8cm}p{11cm}}

$A_{\mathrm{cut},i}$ & Area of the $i^{\mathrm{th}}$ cut.\\
$A_{\mathrm{eff}}$ & Effective cross-sectional area of a bar element.\\
$A_{1},A_{2}$ & Areas of adjacent triangular facets sharing a common edge.\\
$A_{\mathrm{sheet}}$ & Total projected area of the kirigami sheet.\\
$A_{T}$ & Area of the triangular facet associated with a bar element.\\
$c$ & Characteristic distance (effective strip length) between adjacent facet centroids.\\
$\mathbf{C}$ & Bar compatibility matrix.\\
$\mathbf{D}_{B}$ & Diagonal matrix of bending hinge stiffnesses.\\
$\mathbf{D}_{R}$ & Diagonal matrix of rotational spring stiffnesses.\\
$\mathbf{D}_{S}$ & Diagonal matrix of bar stiffnesses.\\
$E$ & Young's modulus.\\
$\mathbf{e}_{S}$ & Vector of bar extensions.\\
$F_{B}$ & Calibration factor for bending hinge stiffness.\\
$F_{S}$ & Calibration factor for bar stiffness.\\
$h$ & Pop-out height induced by buckling.\\
$I$ & Second moment of area.\\
$\mathbf{J}_{B}$ & Jacobian matrix associated with bending hinge rotations.\\
$\mathbf{J}_{R}$ & Jacobian matrix associated with rotational spring elements.\\
$\mathbf{K}_{2D}$ & Global stiffness matrix for the two-dimensional kirigami model.\\
$\mathbf{K}_{3D}$ & Global stiffness matrix for the three-dimensional kirigami model.\\
$\mathbf{K}_{B}$ & Global bending hinge stiffness matrix.\\
$\mathbf{K}_{R}$ & Global rotational spring stiffness matrix.\\
$\mathbf{K}_{S}$ & Global bar stiffness matrix.\\
$K_f$ & Empirical friction coefficient constant.\\
$k$ & Bending curvature.\\
$k_{B}$ & Stiffness of each hinge element.\\
$k_{R}$ & Stiffness of each torsional spring element.\\
$k_{S}$ & Stiffness of each bar element.\\
$k_{\mathrm{sheet}}$ & Equivalent permeability of the kirigami sheet.\\
$l_{\mathrm{cut},i}$ & In-plane length of the $i^{\mathrm{th}}$ cut.\\
$L$ & Bar length or common edge length of adjacent facets.\\
$L_{\mathrm{channel}}$ & Characteristic channel length in the Poiseuille formulation.\\
$L_{\mathrm{Darcy}}$ & Characteristic length in Darcy's law.\\
$M$ & Bending moment.\\
$M_{\mathrm{cont}}$ & Continuum bending moment.\\
$M_{\mathrm{desc}}$ & Discrete bending moment.\\
$N$ & Total number of cuts in the kirigami sheet.\\
$P_{\mathrm{cut},i}$ & Perimeter of the $i^{\mathrm{th}}$ cut.\\
$\Delta P$ & Pressure difference across the kirigami sheet.\\
$Q_{\mathrm{Darcy}}$ & Volumetric flow rate predicted by Darcy's law.\\
$Q_{\mathrm{Poiseuille}}$ & Volumetric flow rate predicted by the Poiseuille model.\\
$S_f$ & Surface factor representing the contact area ratio.\\
$t$ & Thickness of the kirigami sheet.\\
$\mathbf{u}$ & Global nodal displacement vector.\\
$U$ & Strain energy of a rotational spring.\\
$w$ & Effective slit width.\\
$w_i$ & Effective width of the $i^{\mathrm{th}}$ cut.\\

$\alpha$ & Relative angle between two connected bar elements.\\
$\alpha_0$ & Initial (rest) angle of the rotational spring.\\
$\mu$ & Dynamic viscosity of the fluid.\\
$\mu_0$ & Intrinsic base coefficient of friction.\\
$\rho$ & Cut density.\\
$\theta$ & Dihedral rotation angle between adjacent facets.\\

\end{longtable}

\bibliographystyle{unsrt} 
\bibliography{CMAME_refs}


\appendix

\end{document}